\documentclass[nofootinbib,amsfonts,prd,aps,11pt]{revtex4}

\usepackage[T1]{fontenc}
\usepackage{graphicx}
\usepackage{amsmath,amssymb}
\usepackage{hyperref}
\usepackage{graphicx}
\usepackage{subfigure}

\begin{document}

\title{\boldmath Primordial power spectra for scalar perturbations in loop quantum cosmology}

\author{Daniel Mart\'{\i}n de Blas$^{1}$, Javier Olmedo$^{2}$}

\affiliation{1. Departamento de Ciencias F\'isicas, Facultad de Ciencias Exactas, Universidad Andr\'es Bello, Av. Rep\'ublica 220, Santiago 8370134, Chile\\ 2. Department of Physics and Astronomy, Louisiana State University, Baton Rouge, LA 70803-4001, US}

% e-mail addresses: one for each author, in the same order as the authors
%\emailAdd{d.martindeblas@uandresbello.edu}
%\emailAdd{jolmedo@lsu.edu}

\begin{abstract}
  We provide the power spectrum of small scalar perturbations propagating in an inflationary scenario within loop quantum cosmology. We consider the hybrid quantization approach applied to a Friedmann--Robertson--Walker spacetime with flat spatial sections coupled to a massive scalar field. We study the quantum dynamics of scalar perturbations on an effective background within this hybrid approach. We consider in our study adiabatic states of different orders. For them, we find that the hybrid quantization is in good agreement with the predictions of the dressed metric approach. We also propose an initial vacuum state for the perturbations, and compute the primordial and the anisotropy power spectrum in order to qualitatively compare with the current observations of Planck mission. We find that our vacuum state is in good agreement with them, showing a suppression of the power spectrum for large scale anisotropies. We compare with other choices already studied in the literature. 
\end{abstract}

\maketitle

\section{Introduction}

The paradigm of inflation provides nowadays a simple and accurate description of many of the aspects of the universe we observe. It is able to naturally explain several questions like the particle horizon or the flatness problems of early cosmology, among others \cite{liddle}. Remarkably, this paradigm also explains the origin of the large scale structure. It requires, however, the quantum principles for this mechanism to work \cite{lang,muk-brandng}. Then, it is one of the most favorable situations where quantum gravity phenomena can potentially be detected. Regrettably, the traditional paradigms based on quantum field theories on classical spacetimes are not valid close to the Big Bang singularity. They simply assume suitable initial conditions at the onset of inflation or during the slow roll phase, where the only relevant quantum phenomena are those coming from small perturbations of matter and geometry. In this manuscript we are interested in the extension of cosmological perturbation theory to those regimes where classical general relativity is not valid anymore. Among the different candidates for a quantum theory of gravity, Loop Quantum Gravity (LQG) is one of the most developed approaches \cite{thiem}. Loop Quantum Cosmology (LQC), the application of LQG techniques to cosmological scenarios, has demonstrated to be a trustworthy formalism \cite{revLQC}, where the classical singularity is replaced by a quantum bounce, as well as it preserves upon evolution the semiclassicality of quantum states \cite{aps}. This formalism provides an extension of the inflationary scenarios to the Planck era, opening the possibility of potentially studying high energy physics and quantum gravity phenomena. In addition, if inhomogeneities are included, even if one follows the formalism of cosmological perturbation theory, the traditional assumptions adopted there must be revisited if one wants to extend this paradigm of the early universe to the deep Planck regime. For instance, the classical spacetime approximation might not be valid anymore for non-semiclassical states or close to the high curvature regime of the background geometry. In addition, the criteria that pick out an initial state for the inhomogeneities must also be revised in these new quantum gravity scenarios, since these models admit an extension into the high curvature regime, where general relativity breaks down, but the geometry is still regular (though not necessarily classical).

Among the different proposals to study cosmological models that include small inhomogeneities in the framework of LQC so far, there is a well known proposal that adopts the so-called {\it dressed metric} approach \cite{AAN1}. There, the equations of motion of the perturbations evolving on a quantum geometry are influenced by (not all but some of) the fluctuations of the background quantum state. This proposal is inspired by the {\it hybrid quantization} \cite{hybrid}, since it also assumes that the main contributions of quantum geometry are incorporated in the background degrees of freedom, while the inhomogeneities are described by means of a standard representation (for instance a Fock quantization). This is a regime between full quantum gravity and quantum field theories on curved spacetimes. The hybrid quantization was originally applied to Gowdy cosmologies, providing a consistent and successful quantization free of singularities. Afterwards, this quantization was extended to other models, like Gowdy cosmologies coupled to matter \cite{LRS-gowdy}. It was also adopted for the study of cosmological inflationary models with small scalar inhomogeneities, where it was provided a full quantization \cite{hybr-inf1,hybr-inf1b}. However, the dynamics has not been fully understood yet. In Ref. \cite{hybr-inf2} the quantum dynamics was partially solved, assuming a Born--Oppenheimer ansatz for the solutions to the homogeneous scalar constraint, allowing one to recover a dressed metric regime (under certain approximations) from a more general formalism. Finally, in Ref. \cite{hybr-inf3}, a generally covariant formalism for scalar perturbations has been developed in order to strengthen the predictions of this hybrid approach. There, after a canonical transformation in the phase space (of the perturbed theory), the authors are able to explicitly identify the true physical degrees of freedom (Mukhanov--Sasaki variables) without carrying out any gauge fixing. They adopt a Born--Oppenheimer ansatz, derive the dressed metric regime within this fully covariant formalism and provide the effective equations of motion. 

The genuine (loop) quantum dynamics of the background spacetime is a key issue for the dressed metric approach, since all its richness lays in the quantum fluctuations. It is very well known for a free massless scalar field \cite{aps}, but, whenever a potential is added, the dynamics becomes sufficiently intricate that only an effective description has been considered so far \cite{sloan}. Nevertheless, one can restrict the study to those states for which the dressed metric and the effective dynamics are in agreement (assuming that they exist). Within this approximation, the only relevant magnitude to be provided is the value of the scalar field at the bounce (once its mass has been fixed) for the homogeneous model. For the inhomogeneities, nevertheless, one specifies the initial data by choosing a suitable vacuum state (usually) at that time. However, this is the most important conceptual question to be understood. As we mentioned before, this is a qualitatively different situation with respect to the traditional treatments in which suitable initial data is provided at the onset of inflation or during the slow-roll regime. Here, genuine quantum geometry contributions are negligible and then ignored. Presently, there is no well understanding of what is their relevance in these new scenarios. In previous studies \cite{AAN1, AAN2} in loop quantum cosmology, there have been proposed some candidates, the so-called adiabatic states \cite{adiab-LR,adiab-reg-PF,adiab-reg-AP}, mainly focusing in the fourth order ones. On the one hand, the physical predictions obtained from these adiabatic initial conditions for the perturbations at the bounce seem to be in good agreement with observations whenever the scales for which one obtains large corrections are beyond the observable ones in the cosmic microwave background (CMB). In those situations the main quantum corrections only affect the amplitude of the large scale temperature anisotropies of the CMB, i.e. the range of modes that are either not observable or just entering today the horizon. On the other hand, considering adiabatic states of high order (fourth order or more), a ultraviolet renormalization scheme is available for the stress-energy tensor in order to compute the physical energy density of the perturbations 
\cite{adiab-LR,adiab-reg-PF,adiab-reg-AP}. Nonetheless, it is not completely clear that such family of states (or equivalently initial conditions) is the physically preferred choice, at least for those values of the scalar field at the bounce that do not give primordial power spectra where the scales with strong corrections are at the limit of the range of observability or beyond (usually those that do not produce large number of e-foldings).

The purpose of this manuscript is to study different primordial power spectra for scalar perturbations that can be obtained from the hybrid quantized model commented above \cite{hybr-inf2,hybr-inf3}, and qualitatively confront the obtained physical predictions with other approaches in loop quantum cosmology, with particular attention to the dressed metric approach, and with observations provided by Planck mission \cite{planck, planck-inf}. We will assume that the states of the background are highly peaked on the effective geometry, so that it is in good agreement with the dressed metric regime of the hybrid approach.\footnote{It is worth to mention that the quantum corrections obtained for the dynamics of the scalar perturbations in the hybrid quantization approach are expected to be different from the ones obtained with the dressed metric approach because of the different implementation of the polymeric (and inverse volume) corrections.} This consideration allows us to determine the dynamics of the background and the perturbations by means of a set of differential equations that incorporate quantum geometry corrections. For convenience, we will consider initial data at the bounce, since the background dynamics is fully determined by one parameter: the value of the homogeneous mode of the scalar field. We will consider as initial state of the perturbation the adiabatic states studied in Refs. \cite{AAN2} and \cite{adiab-LR} of 0th, 2nd and 4th order. For them, we found that the primordial power spectrum at the end of inflation agrees with observations if there is sufficient e-foldings, but none of the adiabatic states considered in this manuscript provides a suppression of the power spectrum, but an enhancement, for the observable large scale modes. In addition, in order to face the question of the vacuum state of the perturbations at the bounce, i.e. their initial conditions, we have investigated an alternative prescription based on a genuine algorithm that computes numerically an initial vacuum state in such a way that the time variation of the amplitude of the primordial power spectrum of the Mukhanov--Sasaki variable from the bounce to the end of the kinematically dominated epoch is relieved. This method provides a vacuum state at the bounce that produces a primordial power spectrum at the end of inflation that is in good agreement with the current observations of Planck mission. Remarkably, it is suppressed at large scales, as it seems to be favored by current observations. Furthermore, there is an important contrast with respect to the predictions of the adiabatic states since it does not present the highly oscillating region and averaged enhancement at large scales typical of those states in this scenario. 

This paper is organized as follows. In section \ref{sec:eff-dyn} we specify the (semi)classical setting  and its dynamics by means of a set of effective equations of motion within the hybrid quantization approach. The initial value problem is studied in section \ref{sec:init-state}, where we consider several choices of initial vacuum state for scalar perturbations. There we provide a new constructive method to select a suitable initial vacuum state. We also consider adiabatic states to different order. In section \ref{sec:PS-ATPS} we compute the primordial power spectrum at the end of inflation for these vacuum states. We also provide several examples of the power spectrum of temperature anisotropies for the new vacuum state and compare with observations. We conclude and discuss our results in section \ref{sec:dis-conc}. For the sake of completeness, we include an appendix where the previous constructive method is applied to a particular quantum field theory in a de Sitter space.

\section{Effective dynamics of the hybrid approach}\label{sec:eff-dyn}

The system we will study here is a flat Friedmann--Robertson--Walker spacetime with $T^3$ topology. It is endowed with a spacetime metric characterized by a homogeneous lapse $N_0(t)$ and a scale factor $a(t)$ multiplying the auxiliary three-metric $^0h_{ij}$ of the three-torus. The angular coordinates in this spatial manifold are $\theta_i$ such that $2\pi \theta_i/ l_0 \in S^1$, where $l_0$ is the period in each orthonormal direction (for simplicity the three periods have been chosen to be equal). The auxiliary three-metric $^0h_{ij}$ induces a Laplace-Beltrami operator whose eigenfunctions $\tilde Q_{\vec n,\pm} (\vec\theta)$ and eigenvalues $-\omega_n^2=-4\pi^2\vec n\cdot\vec n/l_0^{2}$ are well known (see e.g. Ref. \cite{hybr-inf1b}). Here $\vec n=(n_1,n_2,n_3)\in\mathbb Z^3$ is any tuple whose first component is a strictly positive integer. 

We will adopt a description of this homogeneous geometry in terms of LQG variables. There, one starts with an su(2)-connection and a densitized triad  as basic canonical variables. However, the connection itself is not well defined in the full quantum theory but holonomies of the connection. Besides, in the context of LQC, we will adhere to the improved dynamics scheme \cite{aps}. This choice is motivated by the fact that such cosmologies bounce whenever the energy density achieves a universal critical value $\rho_c\sim\, 0.41 \rho_{\rm Pl}$ for semiclassical states, where $\rho_{\rm Pl}$ is the Planck energy density. In this situation, the classical homogeneous canonical variables of the geometry are the volume $v=l_0^3 a^3$ and the canonically conjugated variable $\beta$ (which is proportional to the Hubble parameter). These variables satisfy the classical algebra $\{\beta,v\}=4\pi G\gamma$, where $G$ is the Newton constant and $\gamma\simeq0{.}2375$ the Immirzi parameter \cite{thiem}. We will couple to this model a massive scalar field $\phi$ fulfilling, with its canonical momentum, $\{\phi,\pi_\phi\}=1$. 

Following the analysis of Ref. \cite{hybr-inf3}, we will introduce scalar perturbations around the classical homogeneous variables. This was done in a closely related scenario in a seminal paper by Halliwell and Hawking \cite{hh}. The perturbative expansion of the Einstein-Hilbert action of general relativity coupled to a massive scalar field is truncated to second order. The resulting action, in the canonical formalism, can be written in terms of a Hamiltonian that is a linear combination of constraints: the homogeneous mode of the scalar constraint that contains quadratic contributions of the perturbations, one Hamiltonian and one diffeomorphism constraints,  both (local and) linear in the perturbations. These two last constraints generate the so-called gauge transformations of the perturbations. The common strategy followed in the theory of cosmological perturbations is to work with gauge invariant potentials \cite{bardeen}. In this sense, in Ref. \cite{lang2} it was proposed to implement the gauge invariant potentials at the action itself by means of the Hamilton-Jacobi theory. However, the explicit dependence of the Hamiltonian in terms of the non gauge-invariant variables was ignored as well as the background variables were not properly transformed.

These last steps were completed in Ref. \cite{hybr-inf3}. Following it, we consider a canonical transformation involving the whole phase space, in order to separate the gauge invariant Mukhanov--Sasaki potential $V_{\vec n,\epsilon}$, and its conjugated momentum ${\pi}_{V_{\vec{n},\epsilon}}$, with respect to the remaining gauge variables, that will be defined as $(V^{(i)}_{\vec n,\epsilon},\pi_{V^{(i)}_{\vec{n},\epsilon}})$, for $i=1,2$. The details can be found in Ref. \cite{hybr-inf3}. The total Hamiltonian, to second order in the perturbations, will be of the form
\begin{equation}
	H_T=\frac{N_0}{16\pi G}\Big(C_0+\sum_{\vec n,\epsilon} C_2^{\vec n,\epsilon}\Big)+\sum_{\vec n,\epsilon} G_{\vec n,\epsilon}\pi_{V^{(1)}_{\vec{n},\epsilon}}+\sum_{\vec n,\epsilon} K_{\vec n,\epsilon}\pi_{V^{(2)}_{\vec{n},\epsilon}},
\end{equation}
where $G_{\vec n,\epsilon}$ and $K_{\vec n,\epsilon}$ are the Fourier modes of the inhomogeneous Lagrange multipliers associated with the constraints linear in the perturbations of the final action. 

We will then adopt a hybrid quantization approach, combining a loop quantization for the homogeneous connection, and a standard representation for the scalar field and the inhomogeneities \cite{hybr-inf1,hybr-inf1b,hybr-inf2,hybr-inf3}. Since the model involves first class constraints, we will adopt the Dirac quantization approach. The physical states must be annihilated by the operators corresponding to the constraints. In this case, one can easily realize that any physical state must be independent of $V^{(i)}_{\vec n,\epsilon}$, for $i=1,2$. The remaining condition to be imposed is that physical states must be annihilated by the homogeneous mode of the Hamiltonian constraint, i.e. by $\hat H=\hat C_0+\sum_{\vec n,\epsilon} \hat C_2^{\vec n,\epsilon}$. Since we do not know yet how to solve this quantum constraint, we will compute approximate solutions by adopting first a Born--Oppenheimer ansatz for them, and then additional approximations (see Refs. \cite{hybr-inf1b,hybr-inf2,hybr-inf3} for further details) that are expected to be fulfilled for instance by semiclassical states with small dispersions. In addition, among them we choose those ones where the effective equations of motion of the expectation values of the basic observables agree with the effective dynamics of loop quantum cosmology (though we do not know yet if this family of states really exists, it seems natural to assume it does). This effective dynamics has been studied in a similar scenario \cite{aps,eff-lqc} corresponding to a flat FRW spacetime filled with a massless scalar field. The effective equations of motion provided there approximate very well the exact quantum evolution of the expectation values of quantum operators for semiclassical states. Following this motivation, the effective Hamiltonian constraint that we will employ in this manuscript is obtained from the quantum Hamiltonian constraint $\hat H$ by replacing expectation values of products of operators by products of expectation values of such operators.

Under these assumptions, the dynamics is generated by the effective Hamiltonian constraint 
\begin{equation}\label{eq:total-const}
	H(N_0)=\frac{N_0}{16\pi G}\Big(C_0+\sum_{\vec n,\epsilon} C_2^{\vec n,\epsilon}\Big),
\end{equation}
with the unperturbed Hamiltonian constraint
\begin{equation}\label{eq:C0-const}
	C_0=-\frac{6}{\gamma^2}\frac{\Omega^2}{v}
	+8\pi G\left(\frac{1}{v}\pi_\phi^2+v m^2\phi^2\right), \quad \Omega=v\frac{\sin(\sqrt{\Delta}\beta)}{\sqrt{\Delta}}.
\end{equation}
In the last expression, $m$ denotes the mass of the massive scalar field and $\Delta=3/(8\pi \gamma^2 G\rho_c)$ the minimum non-zero eigenvalue of the area operator in LQG. Besides, $C_2^{\vec n,\epsilon}$ is the quadratic contribution of each of the modes
\begin{equation}\label{eq:C2-const2}
	C_2^{\vec n,\epsilon}=\frac{8\pi G}{v^{1/3}}\left(\pi^2_{V_{\vec n,\epsilon}}+E^nV_{\vec n,\epsilon}^2\right),
\end{equation}
with
\begin{equation}\label{eq:Es2}
	E^n =\tilde\omega_n^2+m^2v^{2/3}\left(1-\frac{8\pi G \phi^2}{3} +\frac{16 \pi G \gamma \pi_\phi \phi \Lambda}{\Omega^2}\right)+\frac{4 \pi G}{3v^{4/3}}\left(19 \pi_\phi^2-\frac{24 \pi G \gamma^2  \pi_\phi^4}{\Omega^2}\right),
\end{equation}
where $\tilde\omega_n=l_0\omega_n$ and 
\begin{equation}
	\Lambda=v\frac{\sin(2\sqrt{\Delta}\beta)}{2\sqrt{\Delta}}.
\end{equation}
The above expressions are easily obtained from Refs. \cite{hybr-inf2,hybr-inf3} by using the previously mentioned effective dynamics prescription.

It is worth commenting that, for convenience, one can replace ${\cal H}^{(2)}=\frac{3}{4\pi G \gamma^2}\Omega^2-v^2 m^2\phi^2$ by $\pi_\phi^2$ in \eqref{eq:Es2}. This is in agreement with our approximate effective equations of motion and the perturbative scheme we adopt. We have also neglected corrections of the inverse of the volume operator $[1/v]~\simeq~1/v+{\cal O}(v^{-2})$.The effective equations of motion for the phase space variables are obtained by computing their Poisson brackets with the effective Hamiltonian constraint $H(N_0)$ given in the previous expressions. For the background variables the equations of motion are given by
\begin{subequations}\label{eq:hom-eqs2}
	\begin{eqnarray}
		\dot{\phi} &=&  N_0\frac{\pi_\phi}{v}
		+\frac{ N_0}{2v^{1/3}} \sum_{\vec n,\epsilon} E^n_{\phi}V_{\vec n,\epsilon}^2,\\\label{eq:dot-piphi2}
		\dot{\pi}_\phi &=& -N_0vm^2\phi+\frac{ N_0}{2v^{1/3}} \sum_{\vec n,\epsilon} E^n_{\pi_\phi}V_{\vec n,\epsilon}^2,
		\\\label{eq:dot-vol2}
		\dot v &=& \frac{3}{2} N_0v\frac{\sin(2\sqrt{\Delta}\beta)}{\sqrt{\Delta}\gamma}+\frac{ N_0}{2v^{1/3}}\sum_{\vec n,\epsilon}  E^n_{v}V_{\vec n,\epsilon}^2,\\
		\dot \beta &=& -\frac{3}{2} N_0\frac{\sin^2(\sqrt{\Delta}\beta)}{\Delta\gamma}+2\pi G\gamma  N_0 \left(m^2\phi^2-\frac{\pi_\phi^2}{v^2}\right)- \frac{\gamma}{12}\frac{ N_0}{v}\sum_{\vec n,\epsilon}  C_2^{\vec n,\epsilon} \nonumber\\
		&&+\frac{ N_0}{2v^{1/3}}\sum_{\vec n,\epsilon} E^n_{\beta}V_{\vec n,\epsilon}^2,
	\end{eqnarray}
\end{subequations}
where $E^n_{q}=\{E^n,q\}$, for any phase space variable $q$.
For the sake of completeness we have included backreaction contributions, but for all practical purposes we will neglect them in the following. As a consequence, the effective equations of motion for the background will be equivalent to the ones used in Ref. \cite{AAN2} ---see Eqs. (2.9), (2.11) and (4.2)---. The time evolution of the perturbations, on the other hand, is dictated by an infinite number of first order differential equations:
\begin{align}\label{eq:lon-g-pert}
	\dot{V}_{\vec n,\epsilon} =\frac{N_0}{v^{1/3}}\pi_{V_{\vec n,\epsilon}},\quad \dot{\pi}_{V_{\vec n,\epsilon}} =-\frac{N_0}{v^{1/3}} E^n V_{\vec n,\epsilon},
\end{align}
which do not mix different modes and (given our perturbative truncation) are linear in the inhomogeneities. In the following, we will only consider conformal time $\eta$, which involves $N_0=v^{1/3}$. In addition, and for the sake of simplicity, we will set $\omega_n=k$ but with $k$ taking continuous values. This is in agreement with the approximation $l_0\to\infty$. It is enough for all practical purposes to consider $l_0$ much bigger than the Hubble radius during the whole evolution.

\section{Initial state for the Mukhanov--Sasaki scalar perturbations}\label{sec:init-state}

In standard inflationary cosmology it is common to give initial data at the onset of inflation for both the background and the perturbations. One can also consider initial data close to the classical singularity, but keeping in mind that Einstein's theory breaks down there and any physical prediction cannot be fully trusted. Nevertheless, in that regime, one expects that quantum gravity effects will become relevant, and the emergent preinflationary scenario can potentially change the traditional picture. This is indeed the case of our model. In the present scenario the classical singularity is replaced by a quantum bounce whenever the energy density of the system reaches a critical value $\rho_c$ of the order of the Planck energy density.  In LQC usually one considers initial data when the energy density reaches $\rho_c$. This is so because there the Hubble parameter vanishes, the value of the scale factor is chosen equal to the unit (its concrete value is in fact irrelevant for open flat topology), and the momentum conjugated to the field is determined by the scalar constraint (i.e. the Friedmann equation). Therefore, any solution to the effective equations of motion is completely determined by the value of the homogeneous mode of the scalar field at the bounce, i.e. $\phi_B$. In conclusion, concerning initial conditions of the background variables, we consider $\phi_{B}$ as the only free dynamical parameter. In addition, one can also allow for different values for the mass of the scalar field $m$, and then consider it as an additional (homogeneous) free parameter.

However, the specification of initial data for the perturbations is not fully understood yet, though there are several natural candidates as initial state of the inhomogeneous sector. 
This freedom in the choice of initial vacua is tantamount to the fact that quantum field theories in general curved spacetimes do not possess a sufficient number of symmetries allowing to choose a unique vacuum state \cite{uniq-pert, uniq-t3}. The best one can do, by now, is to assume additional physical or mathematical criteria that select a given candidate among all possible choices. 
But let us first remind that the selection of a Fock vacuum state is equivalent to select a complete set of creation and annihilation variables. In turn, this is equivalent to define a complete set of ``positive frequency'' complex solutions $\{v_{k}\}$ to the equation of motion \eqref{eq:lon-g-pert} that satisfy the normalization relation
\begin{equation}\label{eq:norm-cond}
	v_k(v'_k)^*-(v_k)^*v'_k=i.
\end{equation}
In the last equation the prime stands for derivation with respect to conformal time whereas the asterisk stands for complex conjugation. Since the equations of motion for the Mukhanov--Sasaki variables \eqref{eq:lon-g-pert} are linear, the different choices of sets of ``positive frequency'' solutions can be translated to different choices of initial conditions $\{v_{k,0}, v'_{k,0}\}$ at the considered initial time $\eta_{0}$. Also the normalization relation is preserved upon evolution and therefore it is only necessary that such condition is satisfied initially. General initial conditions (up to a irrelevant global phase) can be written as
\begin{equation}
	\label{eq:gen-ini-cond}
	v_{k,0} = \frac{1}{\sqrt{2D_{k}}}, \quad v'_{k,0}= \sqrt{\frac{D_{k}}{2}}\left(C_{k}-i\right),
\end{equation}
where $D_{k}$ is a non-negative function of the mode, $k$, whereas $C_{k}$ is an arbitrary real function. One can restrict the ultraviolet behavior of those functions using well motivated physical and mathematical conditions. For instance, for compact spatial slices one can impose unitary  evolution (in addition to invariance under spatial symmetries) to the Fock quantization \cite{uniq-t3}. Extending this criterion to open spatial sections, for instance by choosing a vacuum with a finite particle production per spatial volume upon evolution, 
one can see that the previous functions must behave like $D_{k} \sim k + \mathcal{O}\left(k^{-\frac{1}{2}-\delta}\right)$ and $C_{k}\sim \mathcal{O}\left(k^{-\frac{3}{2}-\delta}\right)$ for large $k$, with $\delta>0$. Other well known and widely used criteria to select suitable vacuum states, as the Hadamard condition \cite{wald} or the adiabatic states \cite{adiab-LR,AAN2}, give a similar ultraviolet restriction \cite{bib:unitary-de-Sitter}. We will restrict our study to those initial conditions satisfying the unitary evolution requirement. Nonetheless, since this requirement only constraints the ultraviolet behavior, there still exists an infinite freedom. We then need to further restrict the possible candidates. One of the most popular ways to constraint or select a set of ``positive frequency'' solutions (or initial conditions) is by means of the so-called adiabatic states. In order to define a complete set of complex solutions for them one considers the following form\footnote{Note that, since we are working with Mukhanov--Sasaki variables that contains a scaling with respect to the scalar test fields for which the adiabatic solutions where originally defined, the form of the solutions is slightly different \cite{adiab-reg-AP}.}
\begin{equation}
	\label{eq:adiab-sol}
	v_{k} = \frac{1}{\sqrt{2W_{k}(\eta)}}e^{-i\int^{\eta}W_{k}(\bar{\eta})d\,\bar{\eta}}.
\end{equation}
If this expression is introduced in the equations of motion \eqref{eq:lon-g-pert} one obtains for the function $W_{k}$ the following equation 
\begin{equation}
	\label{eq:adiab-sol-W}
	W^{2}_{k} = k^{2} + s - \frac{1}{2}\frac{W''_{k}}{W_{k}}+\frac{3}{4}\left(\frac{W'_{k}}{W_{k}}\right)^{2}.
\end{equation}
Here, $s=s(\eta)$ stands for the corresponding time-dependent mass term which in our case is given by
\begin{equation}
	s = m^2v^{2/3}\left(1-\frac{8\pi G \phi^2}{3} +\frac{16 \pi G \gamma \pi_\phi \phi \Lambda}{\Omega^2}\right)+\frac{4 \pi G}{3v^{4/3}}\left(19 \pi_\phi^2-\frac{24 \pi G \gamma^2  \pi_\phi^4}{\Omega^2}\right).
\end{equation}
Adiabatic solutions are then obtained by choosing conveniently functions $W^{(\mathfrak{n})}_{k}$ that provide approximations to the exact solution to equation \eqref{eq:adiab-sol-W} that converge to them at least as $\mathcal{O}\left(k^{-\frac{1}{2}-\mathfrak{n}}\right)$ in the limit of infinite large $k$, with $\mathfrak{n}$ being the \emph{order} of the adiabatic solution\footnote{The original definition of adiabatic solutions involves an adiabatic time parameter $T$ where the different approximate solutions, i.e. the adiabatic orders, are determined in the limit $T \rightarrow \infty$. Nonetheless, it is well known that this is essentially equivalent to consider the limit of infinitely large $k$.}. It is important to remark that those adiabatic solutions are suitable approximated solutions in the limit of large $k$ but it is also well known that they do not provide in general a good approximation for small $k$. They were originally introduced with the purpose of adopting analytical tools in several situations, and later on it was shown they are also useful for other purposes like renormalization of the stress-energy tensor in cosmological scenarios. However, we are not interested in analytical approximations to the exact solutions but instead in using these adiabatic states to obtain suitable initial conditions for our equations of motion, as it was done in Ref. \cite{AAN2} for the dressed metric approach, in order to study the physical predictions of the hybrid quantization approach, in particular, as well as the robustness of loop quantum cosmology, in general. Actually, from equation \eqref{eq:adiab-sol} one can define initial conditions associated to an adiabatic solution (absorbing an irrelevant global phase) by choosing $D_{k} = W_{k}$ and $C_{k} = -W^{\prime}_{k}/{(2W^{2}_{k})}$, with both expressions in the right hand sides evaluated at $\eta = \eta_{0}$.

Now, we are going to introduce two different procedures that we will use in this work to obtain specific adiabatic initial conditions of different orders. The first procedure mimics the one explained in the reference \cite{adiab-LR}. One obtains an adiabatic solution of order $(\mathfrak{n} + 2)$, $W^{(\mathfrak{n}+2)}_{k}$, by plugging in the right hand side of the equation \eqref{eq:adiab-sol-W} the $(\mathfrak{n})$ order solution $W^{(\mathfrak{n})}_{k}$, starting with $W^{(0)}_{k} = k$ (the solution naturally associated to a free massless scalar field). The other procedure that we use to select specific adiabatic initial conditions for each order is to consider the above mentioned solutions and perform an asymptotic expansion in the limit of large $k$ and then truncating it at the considered order. This method is equivalent to the one used in Ref. \cite{AAN2}. The functions obtained by means of this last procedure will be denoted as $\mathfrak{W}^{(\mathfrak{n})}_{k}$. In the following we will consider these two specific adiabatic initial conditions until 4th order. To show some instances of the functions considered here, obviously both cases lead to the same initial conditions for the 0th order, $W^{(0)}_{k} = \mathfrak{W}^{(0)}_{k} = k$, whereas for the 2nd order we have $W^{(2)}_{k}=\sqrt{k^{2}+s}$ and $\mathfrak{W}^{(2)}_{k} = k + \frac{s}{2k}$. It is worth noting that it is not guaranteed that these two procedures give meaningful initial conditions (and therefore solutions) since they can lead to functions $W^{(\mathfrak{n})}_{k}$ that are negative or even complex for some values of $k$. 

In addition to the above mentioned potential problems of the two procedures to obtain initial conditions associated to adiabatic states it is easy to realize that, in general, both of them give different sets of ``positive frequency'' solutions when considering different initial times. This may not be seen as a big drawback but it is important to stress that, for instance, in de Sitter spacetimes the privileged Bunch--Davies vacuum is selected by one specific set of solutions irrespectively to the time selected to define the corresponding initial conditions. In summary, these adiabatic conditions allow us to reduce the possible initial states for the inhomogeneities, but from the definition of adiabatic states, it is clear that there still exist infinitely many choices of adiabatic states of any order. Recently, it has been proposed a new criterion to select a unique set of initial conditions in the context of adiabatic vacua \cite{pref-inst-vac-AAN}. The criterion selects those initial conditions for which the expectation value of the adiabatically regularized stress-energy tensor at the initial time vanishes.

Besides the previous prescriptions for the choice of initial data, in this work we will also propose a criterion to select a set of initial conditions, but based on different arguments. It focuses on the behavior of the relevant quantity in our computations: the primordial power spectrum of the comoving curvature perturbation ${\cal R}$. The primordial power spectrum ${\cal P}_{\cal R}(k)$ is defined from its 2-point function (which is in one-to-one relation with a concrete vacuum state) as \cite{liddle,lang}
\begin{equation}
	\langle\hat{\cal R}(\eta_{\rm end},{\bf x})\hat{\cal R}(\eta_{\rm end},{\bf x}')\rangle=\int d^3{\bf k}e^{i{\bf k}\cdot({\bf x}-{\bf x}')}\frac{{\cal P}_{\cal R}(k)}{4\pi k^3}\bigg|_{\eta=\eta_{\rm end}},
\end{equation}
where $\eta_{\rm end}$ is the conformal time at which the quantities in the expression above are evaluated (usually at the end of inflation). It can be written, in terms of the ``positive frequency'' solutions of the gauge invariant Mukhanov--Sasaki potential, as
\begin{equation}\label{eq:PS}
	{\cal P}_{\cal R}(k) =\frac{k^3}{2\pi^2}\frac{|v_{k}|^2}{z^2}\bigg|_{\eta=\eta_{\rm end}},
\end{equation}
where $z=\frac{\pi_\phi}{Hv^{2/3}}$ is given in terms of the Hubble parameter defined as $H=\frac{\sin(2\sqrt{\Delta}\beta)}{2\sqrt{\Delta}\gamma}$. 

\begin{figure}
	\centering
	\includegraphics[width=0.49\textwidth]{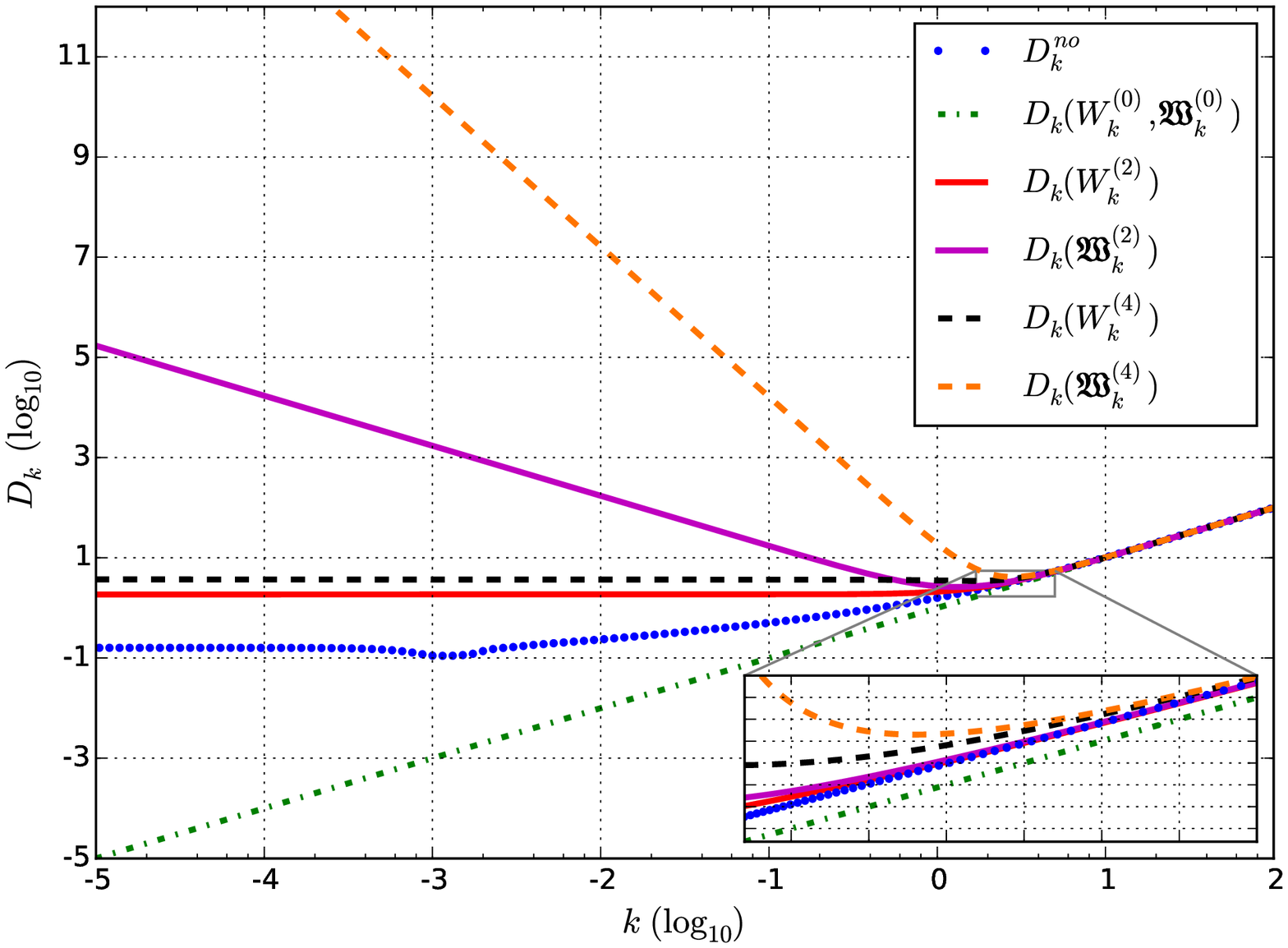} \includegraphics[width=0.49\textwidth]{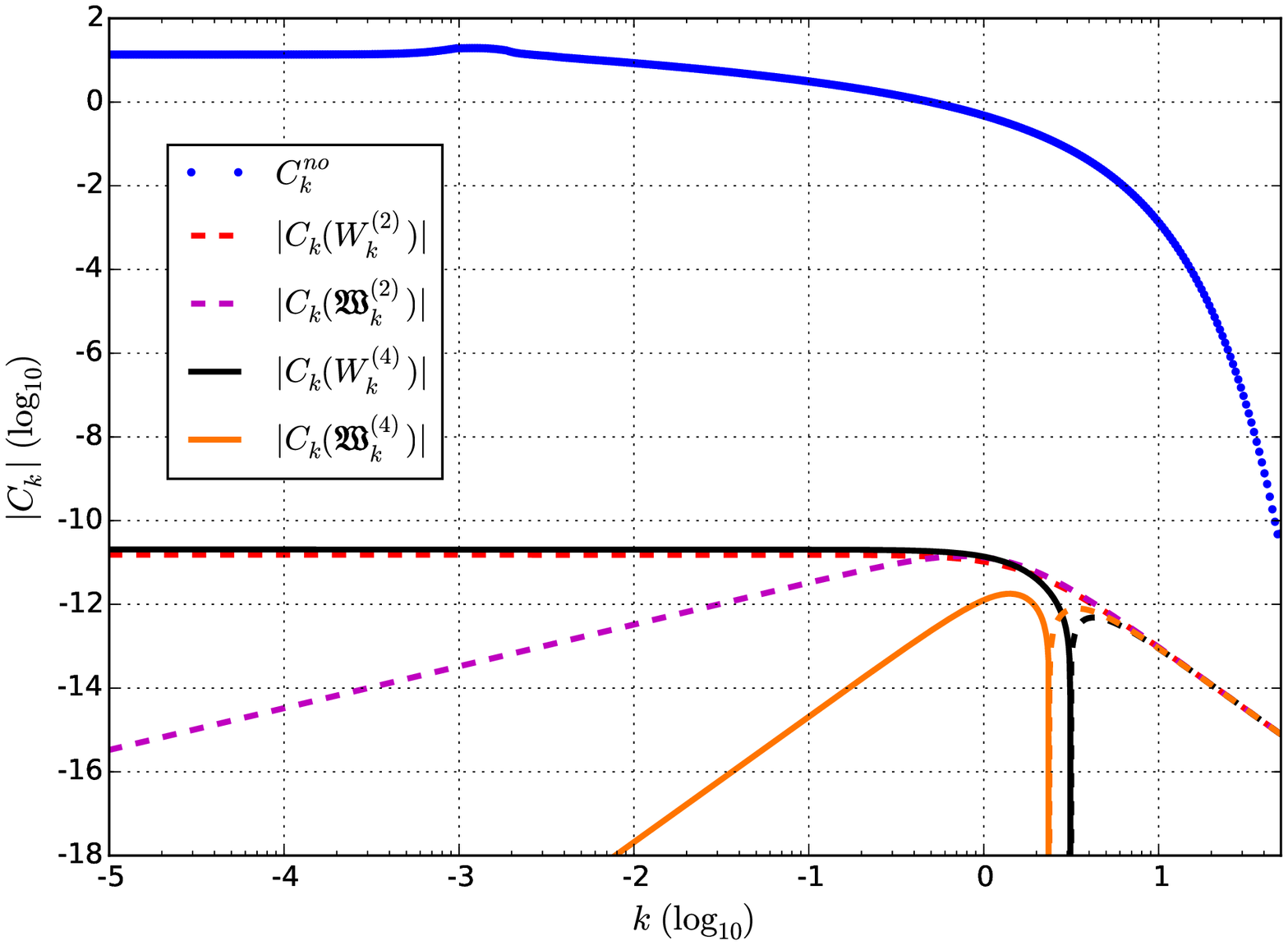}
	\caption{Comparison of the functions that determine the initial condition of the considered vacua for $\phi_{B} = 0.97$ and $m = 1.20\cdot10^{-6}$. Left graph: Comparison of the $D_{k}$ function. Right graph: Comparison of the $C_{k}$ functions. Here the dashed lines correspond to negative values of $C_{k}$. Note that for the 0th order adiabatic initial conditions $C_k(W^{(0)}_{k}) = 0$. }
	\label{fig:Dk-Ck}
\end{figure}

Our criterion is based on selecting the initial conditions that minimize the temporal variation of $|v_{k}|^2$ in a selected time interval. They can be obtained as follows. For each mode, we define the quantity
\begin{equation}\label{eq:IO}
	IO = \int_{\eta_{i}}^{\eta_{f}}\left|\partial_{\bar \eta}|v_{k}|^{2}\right|d\,\bar{\eta},
\end{equation}
that clearly depends on the specific solution $v_{k}$, and therefore on the initial conditions, and also in the considered temporal integration limits. Then, the initial conditions we are looking for are the ones that minimize \eqref{eq:IO}, i.e., we pick out the specific values for the real variables $D_{k}$ and $C_k$ that minimize $IO$. The idea behind this constructive criterion is to select solutions for which the quantity $|v_{k}|^{2}$ does not oscillate rapidly {\it and} with large amplitudes. This choice is justified by the fact that other natural vacuum states in quantum field theories, like the Poincar\'e-invariant vacuum for time-independent scenarios or the Bunch--Davies vacuum for de Sitter spacetimes, are absent of such fast (and large) time-dependent oscillations\footnote{In fact, for time-independent scenarios this criterion selects the Poincar\'e-invariant vacuum and preliminary computations suggest that for de Sitter spacetimes it also obtains the Bunch--Davies vacuum in good approximation whenever the integration time interval is large enough. In App. \ref{app:BD-vac} we show a particular example where we recover the Bunch-Davies state out of this algorithm.}. They are reflected in the corresponding primordial power spectra, showing strong oscillations in $k$. We understand that this behavior should be weakened if one is selecting a more suitable vacuum state (at least approximately). The primordial power spectrum obtained by means of this method is non-highly oscillating and provides the lowest power when averaging in smalls bins of $k$ for $k > 10^{-3}$ (among the states considered in our study). It is worth comenting that, as we show in App. \ref{app:BD-vac}, this criterion allows us to identify the Bunch-Davies vacuum state of a massless scalar field on a de Sitter spacetime. Therefore, we will regard it as a good choice of initial data.
In this work we will take the initial time $\eta_{i}$ as the time of the bounce $\eta_{B}$, which is the one in which we are giving the initial conditions. Furthermore, we set $\eta_{f}$ as the time in which $\phi'$ vanishes for the first time from the bounce, which can be considered as the beginning of the inflationary phase (do not confuse with the slow roll regime). In the following (clearly abusing of the language) we will refer to the initial conditions obtained in this way as the ``non-oscillating'' initial conditions. Nonetheless, it is important to note that, since the time interval considered is finite, this criterion is not expected to give such desirable non-oscillating solutions (in $|v_{k}|^{2}$) for $k \lessapprox 2\pi/(\eta_{f}-\eta_{i})$. In Figure \ref{fig:Dk-Ck} we compare the functions $D_{k}$ and $C_{k}$ for the different adiabatic initial conditions considered here as well as the ``non-oscillating'' ones. As we can see, in the case of $D_{k}$, all choices give the same behavior for $k \gtrapprox 5$, but they strongly differ otherwise. For the functions $C_{k}$ we see that the ``non-oscillating'' initial conditions and the adiabatic vacuum ones differ in the shown range of $k$, although all of them go quickly to zero for large values of $k$.
The ``non-oscillating'' initial conditions, as well as the considered adiabatic initial data, are almost independent of the values chosen for $\phi_{B}$ and $m$ in the ranges that are physically interesting. This is easy to understand taking into account that the actual values of $\phi_{B}$ and $m$ are almost irrelevant in the dynamics of the superinflationary phase after the bounce and the kinematically dominated phase.

\section{Primordial and anisotropy temperature power spectra}\label{sec:PS-ATPS}

\begin{figure}
	\centering
	\includegraphics[width=0.49\textwidth]{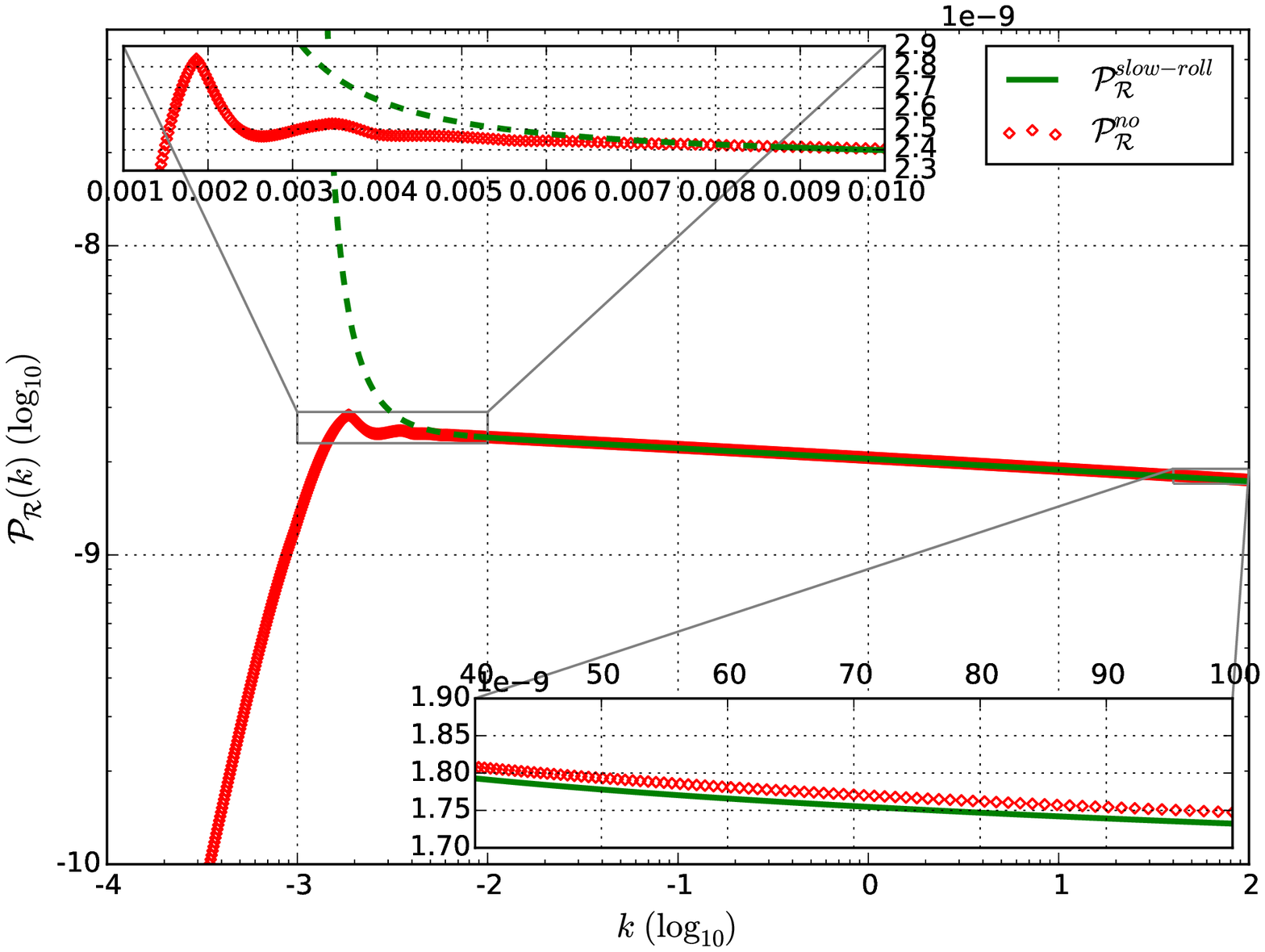} \includegraphics[width=0.49\textwidth]{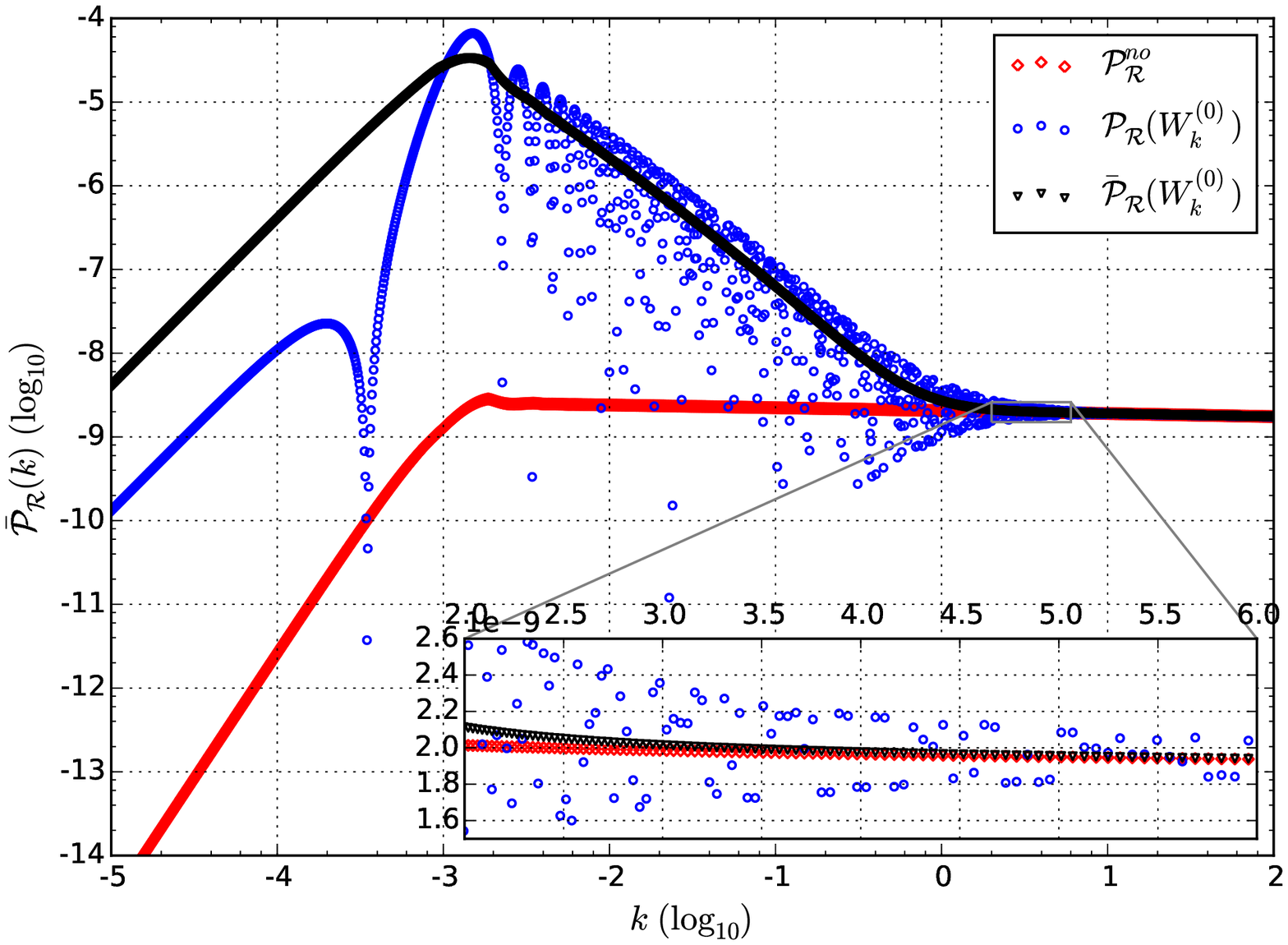}\\
	\includegraphics[width=0.49\textwidth]{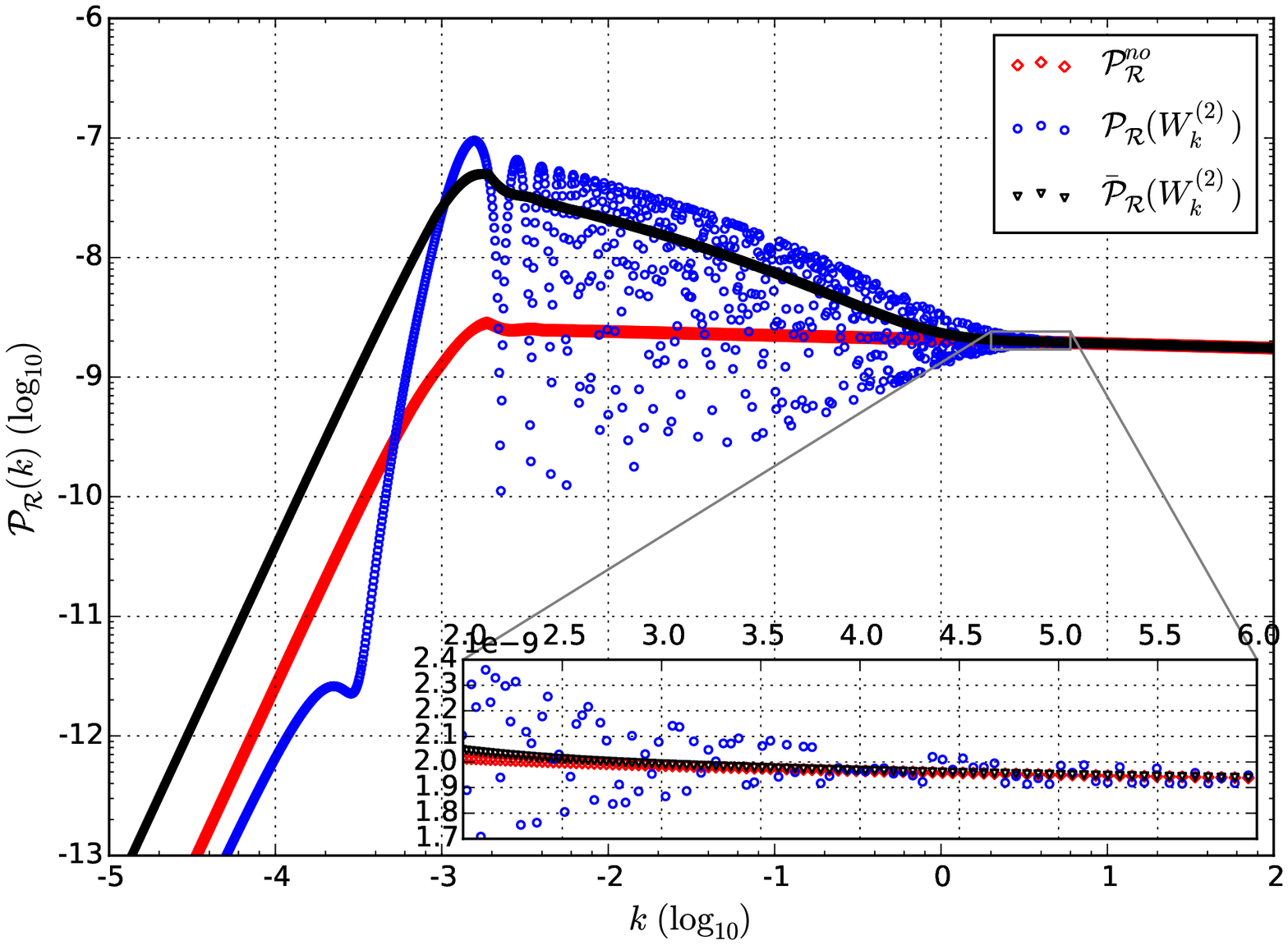} \includegraphics[width=0.49\textwidth]{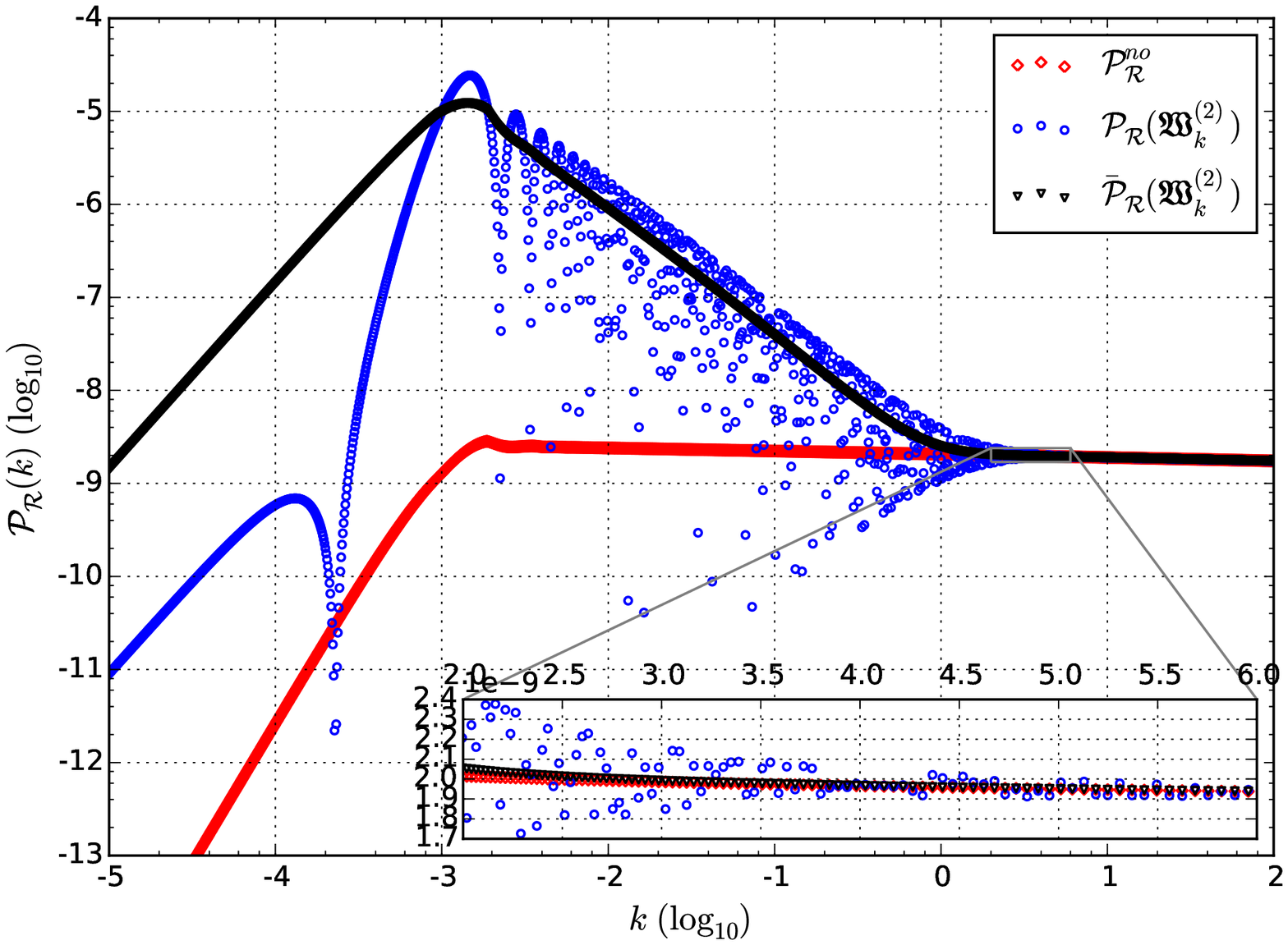}\\
	\includegraphics[width=0.49\textwidth]{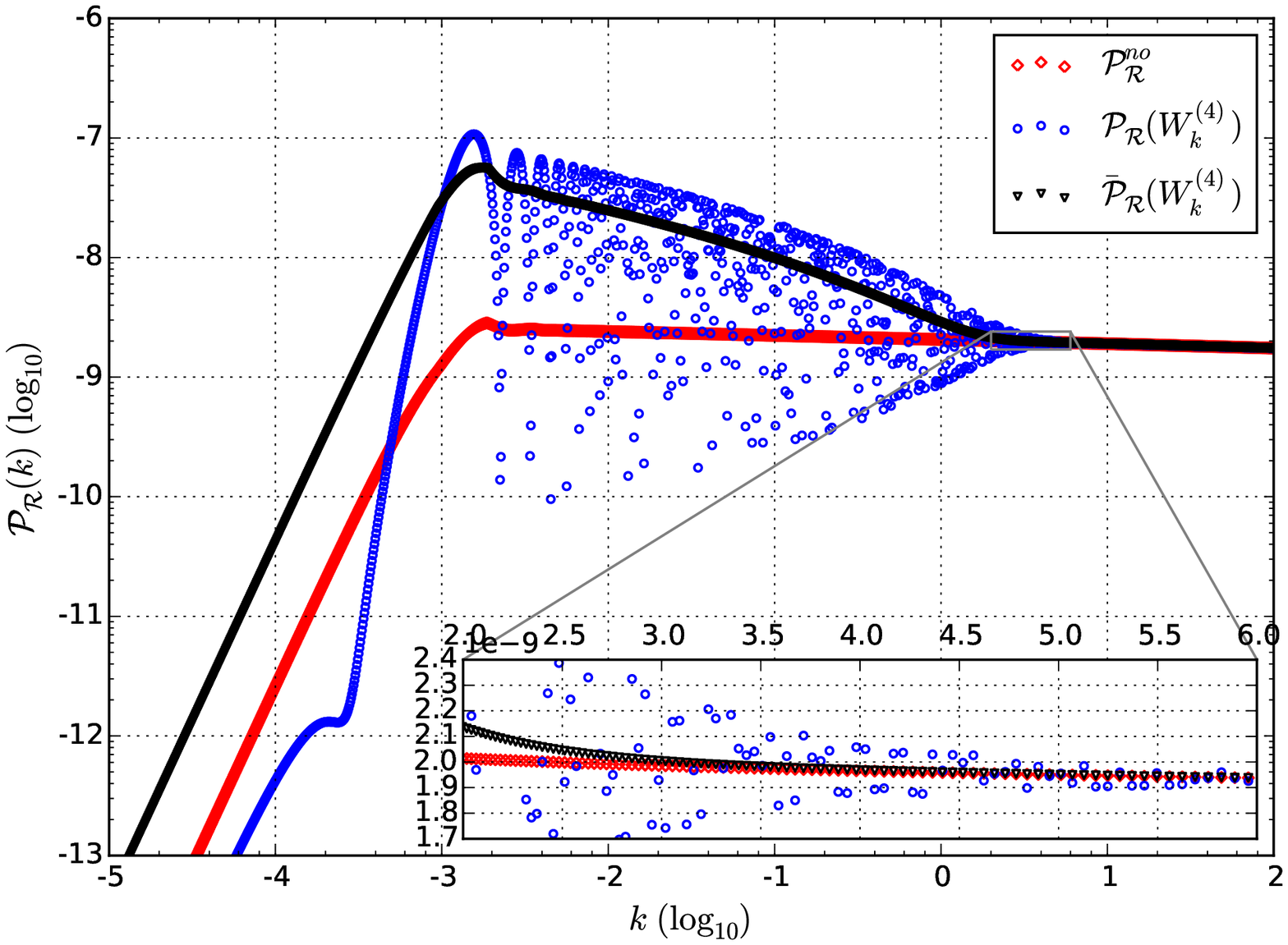} \includegraphics[width=0.49\textwidth]{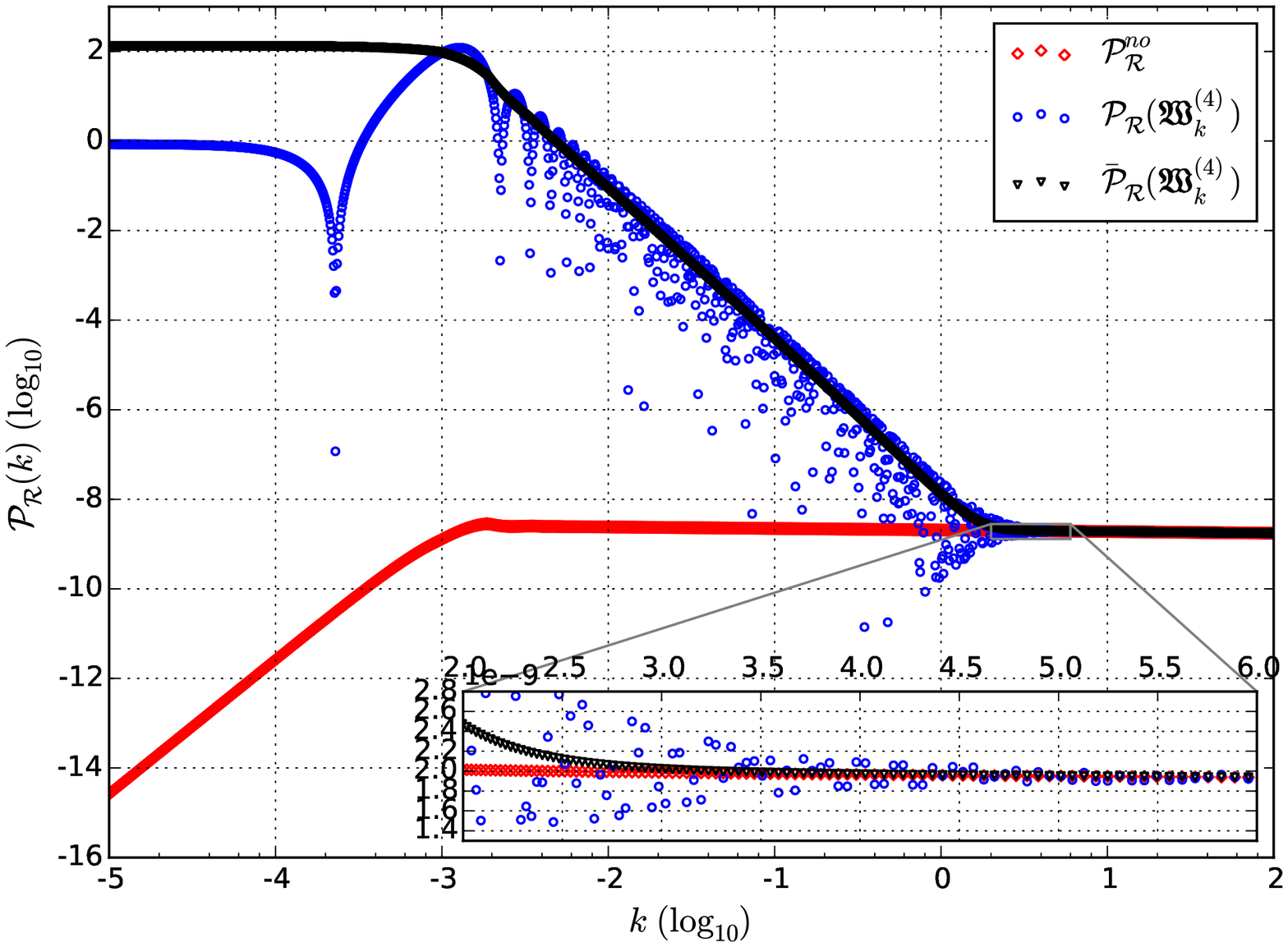}
	\caption{Comparison of the primordial power spectra obtained from the considered sets of solutions for the perturbations and $\phi_{B} = 0.97$ and $m = 1.20\cdot10^{-6}$.}
	\label{fig:PPSs}
\end{figure}

In order to test potential predictions of our proposal regarding the Planck era that can be observed in the cosmic microwave background, we study the primordial power spectrum of the comoving curvature perturbations at the end of inflation ---see equation \eqref{eq:PS}. This magnitude is indirectly related with observations through the angular spectrum of temperature anisotropies of the CMB. For any pair of homogeneous initial conditions, i.e., the mass of the field and value of the field at the bounce, we first compute the primordial power spectrum for the comoving curvature for the ``non-oscillating'' initial conditions, $\mathcal{P}^{no}_{\mathcal{R}}(k)$, which is going to be considered as the reference one. Such primordial power spectrum is obtained from the set of complex solutions $\{v^{no}_{k}\}$, that in the following we are going to denote simply as $\{u_{k}\}$. Any other set of complex solutions is related with the non-oscillatory one by means of
\begin{equation}
	v_{k}=\alpha_{k}u_{k}+\beta_{k}u_{k}^{\ast},
\end{equation}
with $|\alpha_{k}|^2-|\beta_{k}|^2 = 1, \ \forall k$. These mode-dependent complex coefficients $\alpha_{k}$ and $\beta_{k}$ are completely determined by the particular initial conditions $\{v_{k,0},v'_{k,0}\}$ under consideration and the ``non-oscillatory'' ones $\{u_{k,0},u_{k,0}^{\prime}\}$. Actually,
\begin{equation}
	\alpha_{k}=-i\left[(u^{\prime}_{k,0})^{\ast}v_{k,0}-u^{\ast}_{k,0}v^{\prime}_{k,0}\right], \quad \beta_{k}=-i\left[-u^{\prime}_{k,0}v_{k,0}+u_{k,0}v^{\prime}_{k,0}\right].
\end{equation}
Then, one can write the squared absolute value of the new set of complex solutions, and hence the primordial power spectrum, in terms of the reference one as
\begin{equation}
	|v_{k}|^{2}=(1+2|\beta_{k}|^{2}+2\Re[\alpha_{k}\beta_{k}^{\ast}u^{2}_{k}/|u_{k}|^{2}])|u_{k}|^{2},
\end{equation}
where $\Re[\cdot]$ denotes the real part. Since we are considering as the reference set of solutions the ``non-oscillatory'' ones, and taking into account the behavior of the initial conditions considered in the previous section, it is easy to realize that the term producing time oscillations for $|v_{k}|^{2}$ is the one containing the real part of $\alpha_{k}\beta^{\ast}_{k}u^{2}_{k}/|u_{k}|^{2}$. Therefore, for every set of initial conditions we will consider as well the primordial power spectrum obtained by removing that oscillatory part
\begin{equation}
	\bar{\mathcal{P}}_{\mathcal{R}}=(1+2|\beta_{k}|^{2})\mathcal{P}_{\mathcal{R}}^{no}.
\end{equation}
It is worth to comment that for $k \gtrapprox 2\pi/(\eta_{end}-\eta_{B})$, where $(\eta_{end}-\eta_{B})$ denotes the conformal time from the bounce until the end of inflation\footnote{In our simulations we have selected $\eta_{B}=0$. For the instance with $\phi_B = 0.97$ and $m=1.20\cdot 10^{-6}$, the beginning of inflation (defined as the moment in which $\phi'$ equals to zero for the first time) occurs at $\eta_{inf}\approx 2172$, the slow-roll regime starts at $\eta_{onset}\approx2814$ and the inflationary era ends at $\eta_{end}\approx2930$. For other values of $\phi_{B}$ and $m$, in the ranges explored, these times take similar values. Also we have selected the units to perform numerical simulations such that $c = \hbar = G$ = 1.}, $\bar{\mathcal{P}}_{\mathcal{R}}$ will give a good approximation of the averaged power spectrum $\langle\mathcal{P}_{\mathcal{R}}\rangle$ with respect to small bins, whose size in $k$ (or even in $\log k$) must be selected in order average the oscillations (which are expected anyway to be unobservable in the present observations of the CMB). In figure \ref{fig:PPSs}, we compare the primordial power spectra obtained for the initial conditions considered in the previous section, i.e., the ``non-oscillatory'' ones, the free massless initial conditions and the two sets of 2nd and 4th adiabatic initial conditions. Also we compare those primordial power spectra with the one obtained with the first order slow roll formula
\begin{equation}
	\mathcal{P}^{\textrm{slow-roll}}_{\mathcal{R}}=\frac{H^{2}_{\ast}}{\pi \epsilon_{H\ast}},
\end{equation}
where $H$ is the Hubble parameter, $\epsilon_{H} = -\frac{\dot{H}}{H^{2}}$ is the first slow roll parameter in the Hubble flow functions, the subscript $\ast$ means that the quantity is evaluated when the mode crosses  the Hubble horizon $(k = aH)$ and the dot stands for the derivative with respect to the proper time.
As we can see in the first graph of figure \ref{fig:PPSs} the power spectra obtained from the non-oscillatory initial conditions and the slow-roll formula are almost equivalent for modes that cross the Hubble horizon when the slow-roll approximation is valid\footnote{We consider here that slow roll regime starts when the absolute value of both slow roll parameters $\epsilon_{H}$ and $\eta_{H}=\frac{\ddot{H}}{2\dot{H}H}$ is less than $10^{-2}$.}. Nonetheless, we obtain slightly more power in such region for the ``non-oscillatory'' initial conditions because the slow-roll formula is indeed an approximation that is not able (at the considered order) to achieve the required precision. On the other hand, for modes that exit the Hubble horizon before the slow-roll approximation starts to be valid we obtain that the non-oscillating spectrum predicts less power than the slow-roll formula and shows two regions with different behaviors: (i) small oscillations for $ 10^{-3} \lessapprox k \lessapprox 10^{-2}$ and (ii) strong power suppression for $k \lessapprox 10^{-3}$. In the rest of the graphs in figure \ref{fig:PPSs} we show the primordial power spectrum obtained for the adiabatic initial conditions. One can see that for all of them one obtains a highly oscillatory primordial power spectrum. We can distinguish three regions with different behavior. For large $k$ the amplitude of the oscillations tends to vanish and therefore it gets a behavior similar to the ``non-oscillatory'' spectrum and the slow roll formula. For small $k$, one gets a suppression of power for $W^{(0)}_{k}$, $W^{(2)}_{k}$, $\mathfrak{W}^{(2)}_{k}$ and $W^{(4)}_{k}$ and a large constant power for $\mathfrak{W}_{k}^{(4)}$ in the limit $\log k \rightarrow -\infty$. In the intermediate region all the spectra show high oscillations with an averaged enhanced power. It is important to note that these large enhancements are not compatible with current observations unless such region correspond to scales that are not currently observable in the CMB.

\begin{figure}
	\centering
	\includegraphics[width=0.49\textwidth]{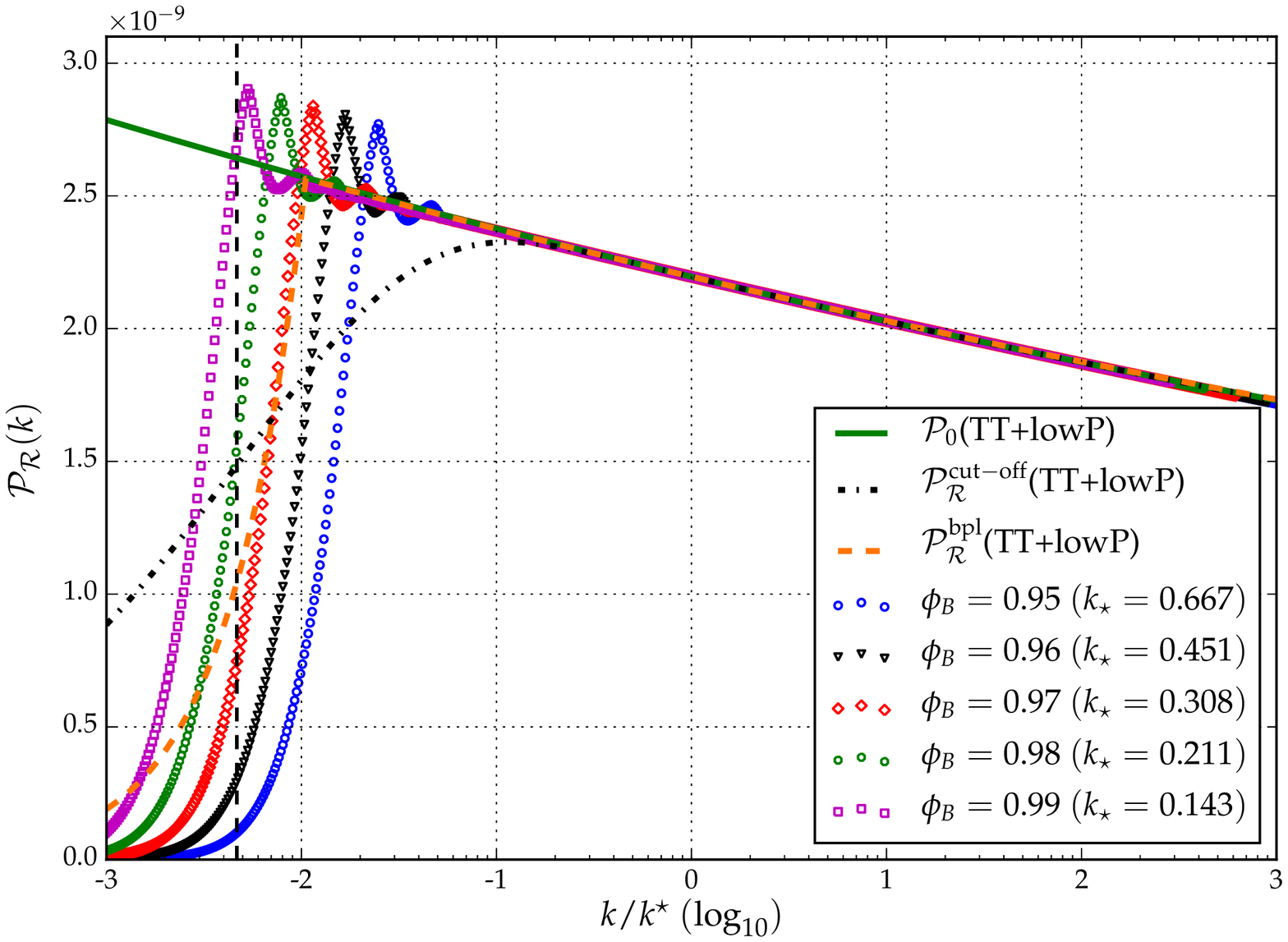} \includegraphics[width=0.49\textwidth]{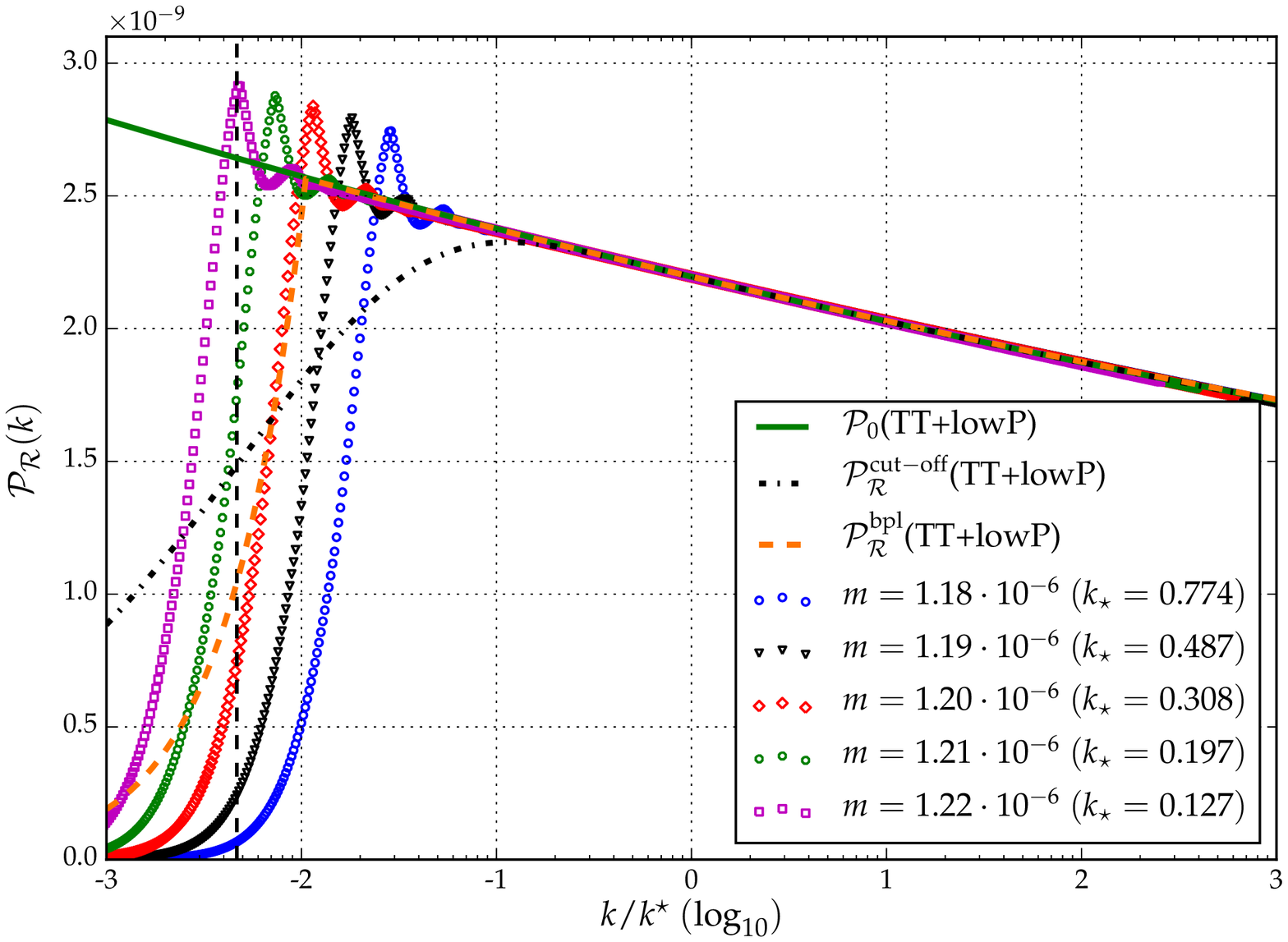}\\
	\caption{Comparison of the primordial power spectra obtained with the ``non-oscillatory'' vacuum for different homogeneous initial conditions with the Planck Collaboration TT+lowP best fit for simple power-law, exponential cut-off and broken-power-law parameterized primordial power spectra. Right graph: Fixed $m = 1.20 \cdot 10^{-6}$. Left graph: Fixed $\phi_{B}= 0.97$. The vertical dashed line corresponds to the largest scale observable in the CMB.}
	\label{fig:PPSskp}
\end{figure}

With the aim at comparing the obtained primordial power spectra with the ones preferred by the observations and statistical analysis of Planck mission one has to perform first a \emph{scale matching}. In fact, we have set arbitrarily\footnote{We must remind that we are using Planck units.} $v_{B} = 1.0$ whereas observationally the scale factor is usually set as the unit nowadays ($v_{o} = 1.0$). Then, in order to obtain a correspondence of the comoving scales $k$ with the (physical) observational ones we (naively) match them using the observational power amplitude $A_{s}$ at some pivot mode $k_{\star}$. We will use here the pivot scale used by the Planck mission \cite{planck-inf} $k_{*} = 0.05\, \textrm{Mpc}^{-1}$ and the amplitude obtained for the best fit\footnote{It is important to note that the amplitude for the best fit depends on the functional form considered for the primordial power spectrum. For the Planck best fit it is considered a simple power-law spectrum ${\cal P}_{\cal R}(k)=A_s(k/k_{\star})^{n_s-1}$.} of the TT+lowP data given by $\log(10^{10}A_{s}) = 3.089\, \pm\, 0.036\, (68\%\, \textrm{CL})$. Therefore, we will define our pivot comoving scale $k_{\star}$ as the one such that $\mathcal{P}^{no}_{\mathcal{R}}(k_{\star})=A_{s}$ and such that it exits the Hubble horizon in (or closer to) the slow-roll region. The value of the pivot scale $k_{\star}$ obtained in this way depends both on the mass of the scalar field $m$ and the value of its homogeneous mode at the bounce $\phi_{B}$. Once such scale matching is done we compare the primordial power spectrum of the ``non-oscillatory'' solutions that we suggest in this manuscript with three parametrized primordial power spectra studied by the Planck Collaboration. The first one is the simple power-law
\begin{equation}
	\mathcal{P}_{\mathcal{R}}(k) = \mathcal{P}_{0}(k) = A_{s}\left(\frac{k}{k_{\star}}\right)^{n_{s}-1},
\end{equation}
which is parametrized by the power amplitude $A_{s}$ at the pivot scale and the spectral index $n_{s}$. The Planck Collaboration obtains from the TT+lowP data that $n_{s} = 0.9655 \, \pm \, 0.0062 \, (68\%\, \textrm{CL})$. This simple form of the primordial power spectrum is the statistically preferred by Planck, not because it provides the best fit to the observational data, but because it yields a good fit with a remarkably small number of parameters. Actually, a better fit to the Planck data is obtained when one allows the primordial power spectrum to deviate from the simple power-law form, either considering a running of the spectral index or different functional forms. Such improvement of the fitting is mainly due to the possibility of suppressing the primordial power spectrum at large scales. In order to obtain a better fit there, the Planck Collaboration considers two forms for the primordial power spectra that include two additional parameters. The first one consists in a simple power-law spectrum multiplied by an exponential cut-off:
\begin{equation}
	\mathcal{P}_{\mathcal{R}}^{\mathrm{cut-off}}(k)=\mathcal{P}_{0}(k)\left\{1-\exp\left[-\left(\frac{k}{k_{c}}\right)^{\lambda_{c}}\right]\right\}.
\end{equation}
This power spectrum is typical in scenarios in which slow roll is preceded by a stage of kinetic energy domination \cite{pow-supp}. The second one is a broken-power-law (bpl) potential of the form
\begin{equation}
	\mathcal{P^{\mathrm{bpl}}_{\mathcal{R}}}=\left\{\begin{array}{ll}
		A_{\textrm{low}}\left(\frac{k}{k_{\star}}\right)^{n_{s}-1 +\delta} & \textrm{if} \ k \leq k_{b},\\
		A_{s}\left(\frac{k}{k_{\star}}\right)^{n_{s}-1} & \textrm{if} \ k \geq k_{b},\\
	\end{array}\right.
\end{equation}
with $A_{\textrm{low}}= A_{s}(k_{b}/k_{\star})^{-\delta}$ to ensure continuity at $k=k_{b}$. The best fit to the TT+lowP Planck data for the first model is given by $\lambda_{c} = 0.50$, $\log(k_{c}/\textrm{Mpc}^{-1}) = -7.98$ and $n_s = 0.9647$. For the second model the best fit is obtained with $n_{s}=0.9658$, $\delta = 1.14$ and $\log(k_{b}/\textrm{Mpc}^{-1}) = -7.55$. Both of them give a slightly better fitting for the observational data, although not good enough to be statistically preferred over the simple power-law spectrum \cite{planck-inf} with two parameters less. In figure \ref{fig:PPSskp} we compare the primordial power spectra for the ``non-oscillatory'' initial vacuum for different values of the homogeneous initial conditions with Planck best fit of the previous parameterized power spectra.
As we see our proposal has a large scale power suppression that resembles the one given by the broken-power-law best fit of Planck for scales observed in the cosmic microwave background. 

\begin{figure}
	\centering     
	\includegraphics[width = 0.49\textwidth]{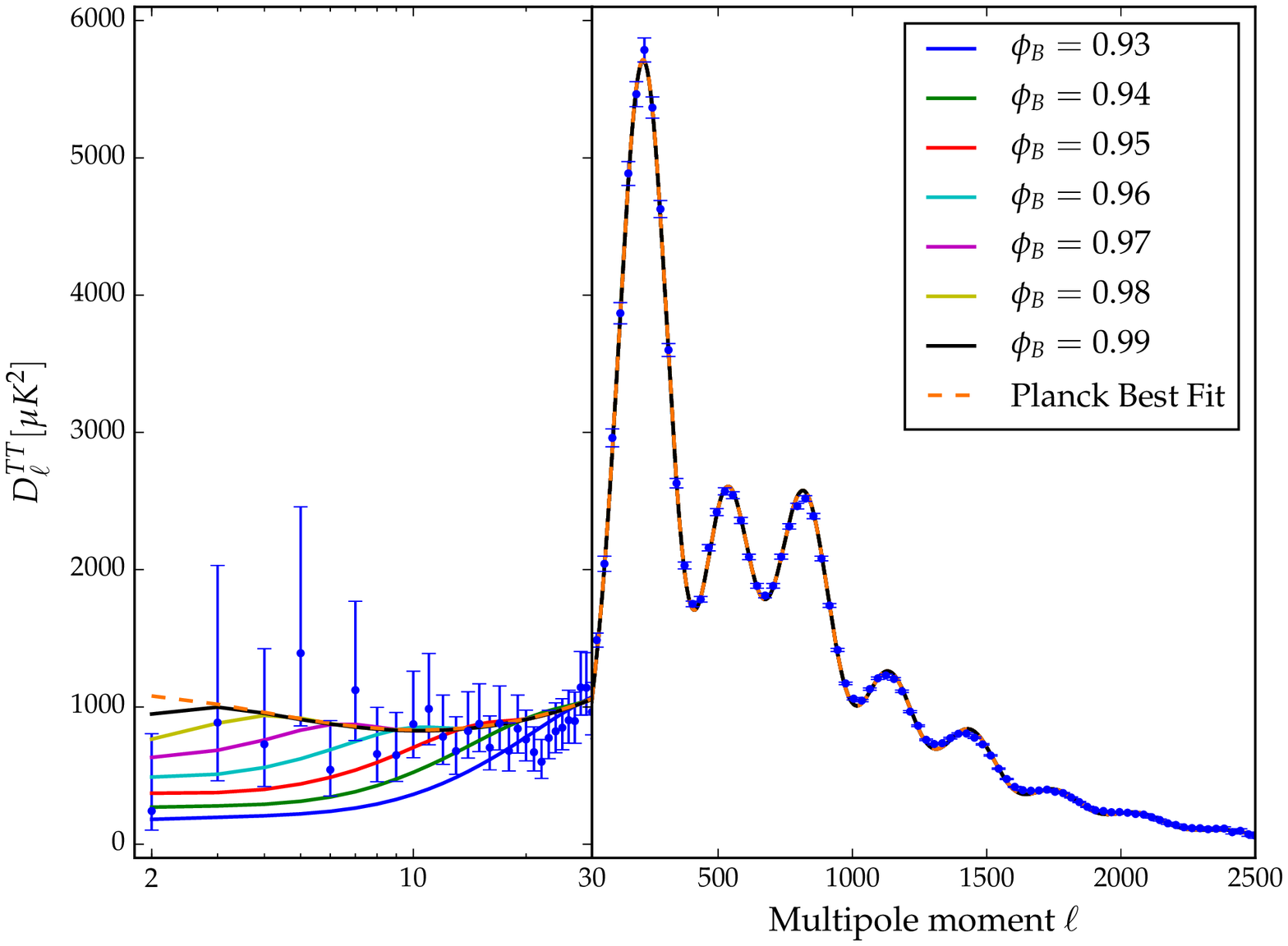}  \includegraphics[width = 0.49\textwidth]{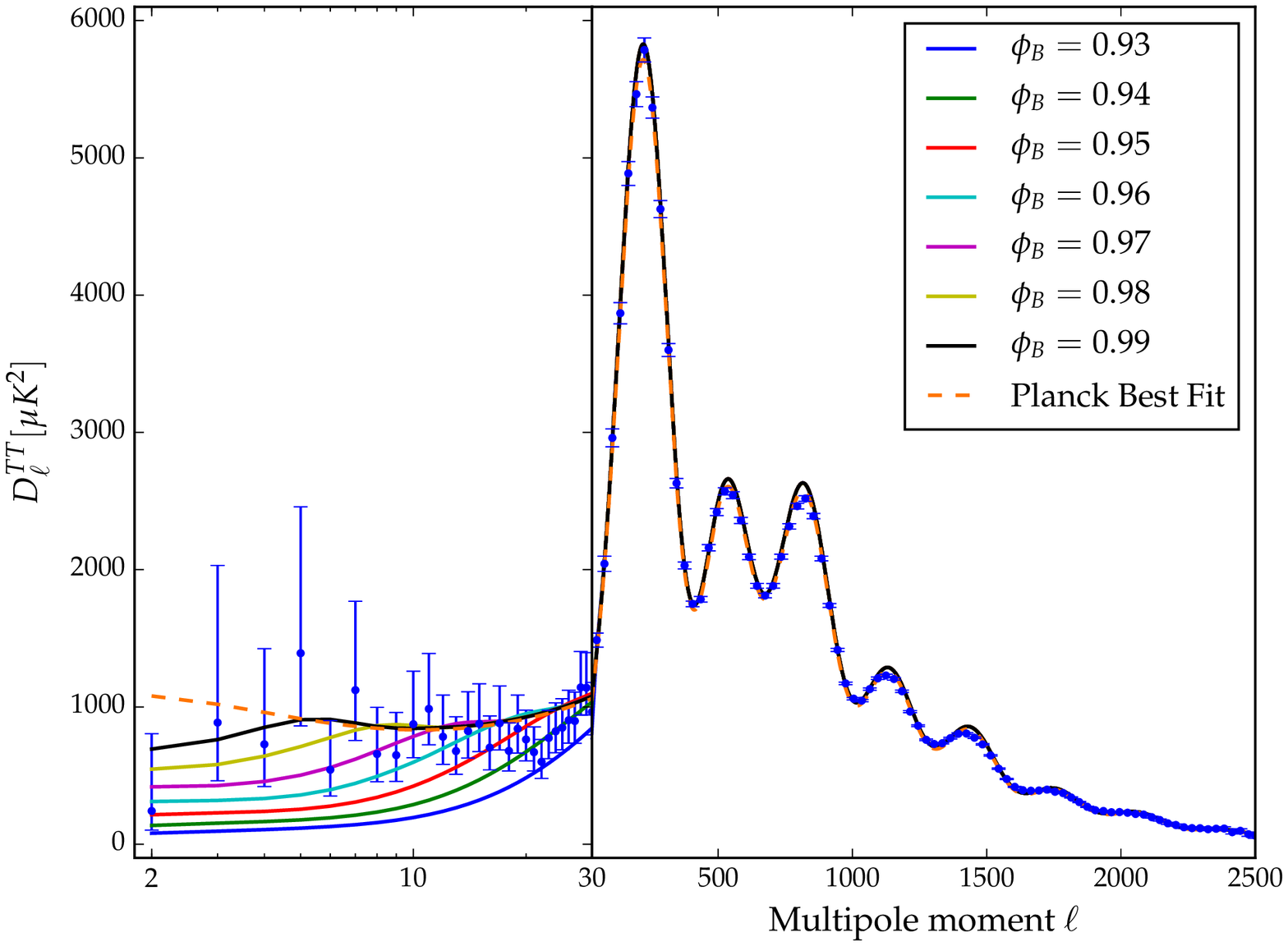}
	\caption{Temperature angular power spectrum provided by Planck best fit and the one computed for the ``non-oscillating'' vacuum state for different values of the scalar field at the bounce and the mass fixed to $m=1.20\cdot10^{-6}$. In the left (right) plot the Planck best fit is computed with a power law (cut-off) primordial spectrum and the corresponding parameters provided by Planck best fit given in Ref. \cite{planck-inf}. }   
	\label{fig:anist}
\end{figure}

Finally, we can see the qualitative consequences of the ``non-oscillatory'' initial conditions at the bounce in the power spectrum of temperature anisotropies. We have carried out the computation employing the \textsf{CLASS} code \cite{class} and the cosmological parameters provided by the Planck Collaboration \cite{planck} for the best fits of both the simple power-law and cut-off models \cite{planck-inf}. In Fig. \ref{fig:anist} we plot the Planck Collaboration observational data for the temperature correlations, the Planck best fit and the predictions of the primordial power spectra obtained for the ``non-oscillating'' initial conditions. We choose a particular value of the mass of the scalar field equal to $m=1.20\cdot10^{-6}$ and several choices of its homogeneous mode at the bounce. We observe that our ``non-oscillating'' initial conditions can explain the power suppression for small angular momenta (large cosmological scales) of the anisotropies of the CMB. This suppression is stronger in those cases where there is not enough inflation, but for a sufficiently high number of e-foldings we still recover the simple power-law primordial power spectrum statistically preferred by Planck. For the left graph in Fig. \ref{fig:anist} we have used the best-fit cosmological parameters provided by the Planck Collaboration whereas for the right graph we use the cosmological parameters for the best fit of cut-off models \cite{planck-inf}. We observe that the peaks of the baryonic resonances are quite sensible to the two sets of cosmological parameters provided by Planck Collaboration, mainly because the values of $\phi_B$ and $m$ must be accurately selected according to the change of the matching scale at the pivot mode, but in both cases they still share the same qualitative properties. Indeed we have not carried out a rigorous statistical analysis in order to determine the best fit parameters for the primordial power spectrum predicted by the ``non-oscillating'' initial conditions. It will be a matter of future research. It is also interesting to notice that it seems that our ``non-oscillating'' vacuum produces a strong suppression of the primordial power spectrum that is not able to explain the anomalies around $\ell=22$, without strongly suppressing the power spectrum of temperature anisotropies at larger scales. Although we have not carried out a detailed analysis, one possible explanation is that the small oscillations that precede the strong suppression of the primordial power spectrum of the ``non-oscillating'' vacuum be responsible of those anomalies  if the amplitude of these oscillations is big enough. However, it is not clear to us by now which physical process could be behind it.

\section{Discussion and conclusions}\label{sec:dis-conc}

In this manuscript we have computed the primordial power spectrum of the comoving curvature perturbation in the framework of hybrid loop quantum cosmology. Since the genuine quantum dynamics has not been fully solved yet, we have considered the effective equations coming from such hybrid quantization (simply replacing operators by expectation values). Additionally, we have neglected the backreaction of the perturbations to the background dynamics. It simplifies the dynamics and allows us to compare the results of the hybrid and the dressed metric approaches. With these assumptions we have computed the primordial power spectrum obtained for different choices of initial vacuum states at the time of the bounce for the Mukhanov--Sasaki variables. More specifically, we have considered two different procedures to obtain specific adiabatic-like initial conditions of arbitrary order and computed the primordial power spectra for 0th, 2nd and 4th orders. They are in good agreement with the ones obtained within the dressed metric approach \cite{AAN2,agu-morr}. Therefore, it is remarkable that the predictions of loop quantum cosmology seem to be robust, since these two formalisms, constructed following different strategies, provide qualitatively similar predictions under the same physical conditions. Note that, although we have neglected backreaction contributions and consider the LQC effective dynamics, the effective equations of motion of the perturbation are different in the hybrid and the dressed metric approaches mainly due to the way in which polymeric corrections are included. The consequence is that the quantitative final results will be different, as well as the adiabatic-like initial data in the two approaches will not agree since they involve time-dependent functions that do not coincide when quantum gravity corrections are important (i.e., at the bounce). The primordial power spectra for adiabatic vacuum states in the hybrid and the dressed metric approaches have in common three distinct behaviors at different scales: (i) smooth slow-roll-like behavior (with small oscillations) for $k \gtrapprox 10$, (ii) large oscillations with an averaged power enhancement for $10^{-3} \lessapprox k \lessapprox 10$ and (iii) strong power suppression for $k \lessapprox 10^{-3}$, except for the expanded adiabatic vacuum of order four $(\mathfrak{W}^{(4)}_{k})$ for which the power remains large and constant, at least in the hybrid approach. It is worth to mention that this kind of primordial power spectra, though they can be in good agreement with observations if the most important corrections correspond to scales that are not currently observable in the CMB, are not able to successfully explain the suppression of the temperature anisotropy power spectrum at large scales, without further considerations \cite{revab,non-gauss}. Let us also comment that, at first glance, neither our results nor the ones of the dressed metric proposal are in agreement with the ones obtained within the \emph{deformed algebra} approach, that leads to highly oscillatory (with large amplitude) primordial power spectra even at small scales \cite{bib:def-al}.

In addition to adiabatic-like initial conditions for the perturbations at the bounce, we have also provided a new criterion to select a suitable initial vacuum state. Such criterion picks out the initial conditions for each mode in such a way that it minimizes the time variation of the amplitude of the Mukhanov--Sasaki variable from the bounce to the beginning of inflation. We have shown that such ``non-oscillating'' initial conditions yield a primordial power spectrum where the large oscillations with the averaged enhanced power are not present, obtaining instead for that range of scales a behavior compatible with the one obtained from the slow-roll formula. Remarkably, it presents a strong power suppression for large scales. Consequently, it may provide a better fitting to the current observations than the spectrum obtained from a quadratic potential considering only the slow-roll regime or from the usual simple power-law primordial power spectrum. Nonetheless, although a rigorous statistical analysis is necessary, it seems that the obtained strong power suppression is not enough to explain the observed anomaly around the multipole $\ell \sim 22$ in the temperature anisotropy angular spectrum. One appealing possibility is to consider initial conditions for the Mukhanov--Sasaki variables that slightly deviate from the ones obtained with the considered criterion. Such initial conditions would lead to a primordial power spectrum with slightly larger oscillations around $k\sim 10^{-3}$ that might explain the above mentioned anomaly. Those slightly differently initial conditions might be obtained by different physical processes. One possibility is, for instance, by minimizing the time integrated (in conformal time) value of the energy density for each mode from the bounce to the beginning of inflation. On the other hand, since we expect that the set of complex solutions selected by ``non-oscillating'' initial conditions give approximately similar physical results to the ones obtained by giving Minkowski-like initial conditions in the kinematically dominated period, another possible way to obtain a primordial power spectrum with larger (but still small) oscillations around $k\sim 10^{-3}$ is defining Minkowski-like initial conditions around the end of the superinflationary era after the bounce. Indeed, it is not clear to us that the quantum dynamics of the Universe is fully determined by our approach but, instead, there is a decoupling scale right after the bounce where the hybrid quantization (and so the dressed metric approach) is valid and where it is natural to give Minkowski-like initial conditions. Finally, another possibility is to break the hypothesis of isotropy before inflation \cite{bianchi1,bianchi2}, being the anisotropies the main source of oscillations at scales of the order of $k\sim 10^{-3}$.

The assumption of isotropy of the universe together with the truncation of the perturbations to second order in the action allow us to focus our attention on scalar perturbations, since they decouple dynamically from the vector and tensor modes. In these circumstances, the vector inhomogeneities are non-dynamical degrees of freedom. However, the tensor modes cannot be neglected if one wants a complete physical picture of the system (under the previous hypotheses). Although the hybrid quantization approach is still incomplete at this respect, our preliminary calculations suggest that it is possible to incorporate tensor modes in this formalism without further considerations. The effective equations of motion are well defined at the Planck regime and have a well behaved ultraviolet limit. Our purpose in the future is to compute the tensor primordial spectra for several initial adiabatic vacuum states (as well as for the "non-oscillatory" one) within this hybrid quantization approach. Although it is soon to draw any conclusion within this formalism, the analyses by Planck Collaboration suggest that a quadratic potential is statistically disfavored since it predicts a tensor-to-scalar ratio slightly higher than the upper bound $r_{0{.}002}<0{.}11$ (95\%CL). However, whether this is true in the hybrid formalism and the possible physical phenomena that could deal with this question is something that we will discuss in a forthcoming publication.

Our results are mainly based on the selection of the ``non-oscillating'' vacuum state, which has been obtained by following a particular algorithm whose main purpose is to minimize the oscillations of the primordial power spectrum. But additional considerations can be further investigated. For instance, the quantity inside the integral in Eq. \eqref{eq:IO} is the absolute value of the derivative of $|v_{k}|^{2}$ with respect to conformal time. Although the algorithm works very well eliminating most of the oscillations in this physical quantity (let us recall that it is related with the 2-point function), one could instead consider minimizing these type of oscillations in other physical quantities like either the total, kinetic or potential time-dependent energy of each mode. Besides, it would be interesting to apply this algorithm to different cosmological settings. In particular, there are some situations where this method can be tested since the natural vacuum state is already known. This is the case, for instance, of time-independent scenarios or de Sitter spacetimes. If we consider arbitrary initial conditions, like in Eq. \eqref{eq:gen-ini-cond}, it would be very interesting to see if this algorithm is able to reconstruct the privileged initial conditions of the Poincar\'e-invariant vacuum or Bunch-Davies vacuum, respectively. A preliminary study for a test, massive scalar field in a Minkowski spacetime shows that the algorithm is able to find in a good approximation the initial conditions for the Poincar\'e-invariant vacuum state. All these aspects will be addressed in future publications.

In summary, the study carried out in this manuscript provides novel ideas about the extension of the traditional inflationary paradigm of cosmological perturbations theory to the Planck era. When the Big Bang singularity is avoided, and the evolution of the Universe can be extended far in the past with respect to the onset of inflation, it is natural to ask again these two important questions: i) what is the natural initial state of the Universe in the past, for instance, at the high curvature regime? and ii) how predictive are these new scenarios with respect to the present observations? We show here that loop quantum cosmology and the hybrid quantization approach of this particular model suggest a possible answer to these questions. Besides, the usually ignored freedom about the choice of initial state of the perturbations has been considered in this manuscript. We have provided a new criterion that can seed light on the understanding about the existence of privileged vacuum states that has not been considered before in the way we do here, at least to the knowledge of the authors. We also strongly believe that our criterion is not restricted to scenarios in genuine quantum cosmology but they can also be adopted in many other models in cosmology without further considerations.

\acknowledgments

The authors are greatly thankful to I. Agull\'o, L. Castell\'o Gomar, M. Mart\'in Benito, G. A. Mena Marug\'an and T. Paw{\l}owski for enlightening conversations and suggestions reflected in the manuscript. We also thank J. Torrado for his clarifications about the \textsf{CLASS} code. This work was supported in part by Pedeciba, and the grants MICINN/MINECO FIS2011-30145-C03-02 and FIS2014-54800-C2-2-P from Spain. D. M-dB is supported by the project CONICYT/FONDECYT/POSTDOCTORADO/3140409 from Chile. J. O. acknowledges support by the grant NSF-PHY-1305000 (USA).

\appendix

\section{Bunch-Davies vacuum and the non-oscillating criterion}\label{app:BD-vac}

In this appendix we will show that the criterion of minimizing (mode by mode) the time variation of the power spectrum, introduced in the section \ref{sec:init-state}, allows us to pick out the Bunch-Davies state when considering a test scalar field in a de Sitter spacetime. We will restrict the study to a massless scalar field in the cosmological chart of this spacetime, as it was done in Ref. \cite{lang}.\footnote{We will study a more general proof (for massive and massless scalar fields on the full de Sitter spacetime) in a future publication.} We will first briefly summarize what is the Bunch--Davies vacuum and how it can be selected by using the symmetries of the de Sitter spacetime and the Hadamard condition. Further details can be found in Refs. \cite{allen,allen-folacci,bib:unitary-de-Sitter}. The action of the test massless scalar field, $\phi$, is given by
\begin{equation}\label{eq:action}
	S=\int d^4x\sqrt{-g}\left(-\frac{1}{2}\partial_\mu\phi\partial^\mu\phi\right),
\end{equation}
where $g$ denotes the determinant of the metric $g_{\mu\nu}$ that in conformal time takes the form
\begin{equation}
	ds^2=a^2(\eta)\left(-d\eta^2+d{\bf x}^2\right).
\end{equation}
Here, $\eta\in(-\infty,0)$ and 
\begin{equation}
	a(\eta)=-\frac{1}{H\eta},
\end{equation}
with $H$ the constant Hubble parameter (in cosmic time). 

It is worth commenting that, due to homogeneity and isotropy of this spacetime, the metric is invariant under rotations and translations. In addition, it is invariant under the dilatations $(\eta,{\bf x})\to (e^{\lambda}\eta,e^{\lambda}{\bf x})$. These transformations leave $H$ unaltered. The consequence is that the previous action in Eq. \eqref{eq:action} will be also invariant under this set of transformations. 

Let us now consider the redefinition $\varphi(\eta,{\bf x})=a(\eta)\phi(\eta,{\bf x})$ and decompose the field $\varphi$ in Fourier modes of the form of $u_{\bf k}(\eta)e^{i{\bf k}\cdot{\bf x}}$. Then, the partial differential field equation can be written in terms of a set of infinitely many ordinary differential equations given by
\begin{equation}\label{eq:desitter}
	u_{\bf k}''(\eta)+\left(k^2-\frac{2}{\eta^2}\right)u_{\bf k}(\eta)=0,
\end{equation}
with $k^2={\bf k}\cdot{\bf k}$. One can easily see that these equations are invariant under the set of transformations considered above, in agreement with the action in Eq. \eqref{eq:action}. 

The solutions to these equations are known,
\begin{equation}\label{eq:sol-ab}
	u_{\bf k}(\eta)=\alpha_{\bf k}\frac{e^{-i k\eta}}{\sqrt{2k}}\left(1-\frac{i}{k\eta}\right)+\beta_{\bf k}\frac{e^{i k\eta}}{\sqrt{2k}}\left(1+\frac{i}{k\eta}\right),
\end{equation}
where $\alpha_{\bf k}$ and $\beta_{\bf k}$ are complex constants, that in principle may depend on the mode ${\bf k}$. Let us remind that the choice of a particular vacuum for the Fock quantization is tantamount to select a complete set of solutions to the equations of motion satisfying the normalization condition given in Eq. \eqref{eq:norm-cond}. The normalization condition imposes $|\alpha_{\bf k}|^2-|\beta_{\bf k}|^2 = 1$ and, taking into account the irrelevance of a complex global phase, the freedom on selecting a vacuum state is given by two real parameters per mode. 
Nonetheless, if one imposes invariance of the vacuum under rotations then the complex constants can only depend on $k$. In addition if one requires invariance under dilations then $\alpha_{\bf k}$ and ${\beta_{\bf k}}$ must be k-independent, therefore $\alpha_{\bf k} = \alpha$ and $\beta_{\bf k} = \beta$. 
The Bunch--Davies vacuum is obtained by taking $\alpha = 1$ (and therefore $\beta = 0$) and it is the unique vacuum which is invariant under the afore mentioned symmetries and its 2-point function has the Hadamard form \cite{allen}. In order to elaborate more about this point, let us consider symmetry invariant vacuum states that we will denote as $|\alpha,\beta \rangle$ and obtain the 2-point function, for simplicity, evaluated at the same time $\eta$. It is given by
\begin{equation}\label{eq:2pointf}
	G^{\varphi}_{\alpha\beta}(\eta;{\bf x},{\bf x}')=\langle\alpha,\beta|\hat\varphi(\eta,{\bf x})\hat\varphi(\eta,{\bf x}')|\alpha,\beta\rangle.
\end{equation}
Let us comment that the 2-point functions in this 2-parameter family are invariant under spatial translations and rotations, and dilations of the space and time. The 2-point functions of the field $\phi$ are easily related with the previous ones by
\begin{equation}
	G^{\phi}_{\alpha\beta}(\eta;{\bf x},{\bf x}')=a^{-2}(\eta)G^{\varphi}_{\alpha\beta}(\eta;{\bf x},{\bf x}').
\end{equation}

One can compute explicitly these 2-point functions \cite{allen}. If we expand the quantum field $\hat\varphi$  as
\begin{equation}
	\hat\varphi(\eta,{\bf x})=\frac{1}{(2\pi)^{3/2}}\int d^3{\bf k}\left(\hat a^{\alpha\beta}_{\bf k}u_k^{\alpha\beta}(\eta)e^{i{\bf k}\cdot{\bf x}}+ \big(\hat a^{\alpha\beta}_{\bf k}\big)^\dagger\big(u_k^{\alpha\beta}(\eta)\big)^*e^{-i{\bf k}\cdot{\bf x}}\right)
\end{equation}
such that $\hat a^{\alpha\beta}_{\bf k}|\alpha,\beta\rangle=0$, the expectation values in Eq. \eqref{eq:2pointf} can be computed and yield
\begin{equation}
	G^{\varphi}_{\alpha\beta}(\eta;{\bf x},{\bf x}')=(|\alpha|^2+|\beta^2|)P_1(\eta;{\bf x},{\bf x}')+\Re(\alpha\beta^*)P_2(\eta;{\bf x},{\bf x}')+i \Im(\alpha\beta^*)P_3(\eta;{\bf x},{\bf x}'),
\end{equation}
where $\Im(\cdot)$ stands for the imaginary part and the functions $P_i$ are defined as
\begin{equation}
	P_i(\eta;{\bf x},{\bf x}')=\frac{1}{(2\pi)^{3}}\int d^3{\bf k}e^{i{\bf k}\cdot({\bf x}-{\bf x}')}Q^{(i)}_k(\eta),
\end{equation}
such that, 
\begin{align}
	Q^{(1)}_k(\eta)&= |v_k(\eta)|^2,\\
	Q^{(2)}_k(\eta)&=v_k(\eta)v_k(\eta)+v^*_k(\eta)v^*_k(\eta),\\
	Q^{(3)}_k(\eta)&=v_k(\eta)v_k(\eta)-v^*_k(\eta)v^*_k(\eta).
\end{align}
Here, we have introduced the function 
\begin{equation}
	v_k(\eta)=\frac{1}{\sqrt{2k}}\left(1-\frac{i}{k\eta}\right)e^{-ik\eta}.
\end{equation}

Then, we immediately see that $Q^{(2)}_k(\eta)$ and $Q^{(3)}_k(\eta)$ oscillate as functions of $\eta$. This property will be essential in our criterion for the choice of vacuum state. The previous integrals can be computed (see Ref. \cite{allen}), and one can obtain explicitly $G^{\varphi}_{\alpha\beta}(\eta;{\bf x},{\bf x}')$. However, we will not give here the explicit result. Nonetheless, one can check in Ref. \cite{allen} that among these 2-point functions, there is only one that is Hadamard \cite{allen}. Precisely, it is the only 2-point function where the contributions of $Q^{(2)}_k(\eta)$ and $Q^{(3)}_k(\eta)$ to the power spectrum disappear since it corresponds to $\alpha=1$ and $\beta=0$. As we mentioned before, this choice corresponds to the so-called Bunch-Davies vacuum state.

As we have shown any vacuum state is obtained from a complete set of complex solutions $\{v_{k}\}$. Therefore, given an initial time $\eta_{0}$, the initial data $\{v_{k,0},v'_{k,0}\}$ that select a set of solutions for the Bunch--Davies vacuum are
\begin{equation}
	v_{k,0} = \frac{1}{\sqrt{2k}}\left(1-\frac{1}{k\eta_{0}}\right)e^{-ik\eta_{0}}, \qquad v'_{k,0} = -i\sqrt{\frac{k}{2}}\left(1-\frac{i}{k\eta_{0}}-\frac{1}{k^{2}\eta^{2}_{0}}\right)e^{-ik\eta_{0}}.
\end{equation}
Taking into account that the addition of a global phase to the solutions still defines the same vacuum state, we will consider instead initial data $\{\tilde{v}_{k,0},\tilde{v}'_{k,0}\}$ of the form given in Eq. \eqref{eq:gen-ini-cond} with
\begin{equation}
	D^{\rm BD}_{k} = \frac{k}{1+\frac{1}{k^{2}\eta^{2}_{0}}}, \qquad C^{\rm BD}_{k}= -\frac{1}{k^{3}\eta_{0}^{3}},
\end{equation}
where the superscripts are referred to the Bunch-Davies vacuum. 
It is straightforward to see that with these initial data one gets the equivalent set of solutions $\{\tilde{v}_{k} = v_{k}e^{i\sigma_{k}}\}$, where $\sigma_{k}~=~k\eta_{0}~+~\arctan(\frac{1}{k\eta_{0}})$.

Now, we will show the consequences of our criterion when it is applied to this particular quantum field theory. We will only assume invariance under translations and rotations in what follows. Let us then start by choosing a set of arbitrary initial data $\{u_{k,0},u'_{k,0}\}$ accordingly, for a given initial time $\eta_0$, of the form of Eq.  \eqref{eq:gen-ini-cond}. One can easily see that the resulting set of solutions $\{u_{k}\}$ can be written as in Eq. \eqref{eq:sol-ab}. This time, the coefficients, in principle, can depend on $k$. Indeed, if $\alpha(k)=|\alpha(k)|e^{i\phi_\alpha(k)}$ and $\beta(k)=|\beta(k)|e^{i\phi_\beta(k)}$, we get 

\begin{align}
	|\alpha(k)|^2& =\frac{1}{2} + \frac{k}{D_{k}\left(1+\frac{1}{k^{2}\eta^{2}_{0}}\right)}\left(1+\frac{1}{k^6\eta^{6}_{0}}\right) + \frac{D_{k}\left(1+\frac{1}{k^{2}\eta^{2}_{0}}\right)}{k}\left(1+C^{2}_{k}\right)+ \frac{2C_{k}}{k^{3}\eta^{3}_{0}} ,\\
	|\beta(k)|^2&=|\alpha(k)|^2-1,\\
	\tan\phi_\alpha(k)&=\frac{1 + C_k D_k \eta_0 + C_k D_k \eta_0^3 k^2}{D_k \eta_0 + D_k \eta_0^3 k^2 + \eta_0^3 k^3},\\
	\tan\phi_\beta(k)&=\frac{1 + C_k D_k \eta_0 + C_k D_k \eta_0^3 k^2}{D_k \eta_0 + D_k \eta_0^3 k^2 - \eta_0^3 k^3}.
\end{align}
We will now concentrate only in the Fourier transform of the two point function evaluated at the same time $\eta$. It is essentially given by Eq. \eqref{eq:PS} as
\begin{equation}
	{\cal P}_{\varphi}(k) =\frac{k^3}{2\pi^2}|u_{k}(\eta)|^2,
\end{equation}
with
\begin{equation}
	|u_{k}(\eta)|^2 = \left[|\alpha(k)|^2 + |\beta(k)|^2 + 2 |\beta(k)| |\alpha(k)| \cos\left(\Phi_k(\eta)\right)\right]\frac{1}{2k}\left(1+\frac{1}{k^2\eta^2}\right),
\end{equation}
and\begin{equation}
	\Phi_k(\eta)=-2 k (\eta - \eta_0) + \phi_\alpha(k) - \phi_\beta(k) + 2 \arctan\left[-\frac{1}{k \eta}\right] - 2 \arctan\left[-\frac{1}{k \eta_0}\right].
\end{equation}
These are the basic ingredients we need in order to proceed with our method. We should minimize the integral in Eq. \eqref{eq:IO} with respect to $D_k$ and $C_k$.
Numerically, it is possible to minimize this function for any choice of ${\eta_{i}}$ and ${\eta_{f}}$. Of course, for those modes that have not oscillated several times in the interval $[{\eta_{i}},{\eta_{f}}]$, this algorithm will not yield an accurate value of the minimum. This is why we must consider $[{\eta_{i}},{\eta_{f}}]$ to be sufficiently extensive, covering a large portion of the spacetime.

Since it is computationally expensive, we will consider instead the minimization of the integral
\begin{equation}\label{eq:IO2}
	IO_2 = \int_{\eta_{i}}^{\eta_{f}}\left(\partial_{\bar \eta}|u_{k}|^{2}\right)^2d\,\bar{\eta},
\end{equation}
which yields the same result, at least for conformal time,\footnote{Notice that Eq. \eqref{eq:IO} is invariant under time reparametrizations, while Eq. \eqref{eq:IO2} is not} but with this last expression easier to handle analytically than Eq. \eqref{eq:IO}. The integral can be computed explicitly. Nevertheless, we will not show here the result since the expression is considerably long. Indeed, it is enough to consider, for any fixed value of $\eta_{f}$, the behavior of the integral when $\eta_{i}\to-\infty$. It is possible to see that it diverges as
\begin{equation}
	IO_2\simeq 2 \eta_i (|\beta(k)|^2 + |\beta(k)|^4) + \mathcal{O}\!\left(1\right),
\end{equation}
for any $k$ if $|\beta(k)| \neq 0$. Here we have used $|\alpha(k)|^2=1+|\beta(k)|^2$. Nevertheless, if $|\beta(k)|=0$ for all $k$, the integral always converges, since it behaves as 
\begin{equation}
	IO_2=\frac{2}{3k^6}\left[\frac{1}{\eta_i^5}-\frac{1}{\eta_f^5}\right].
\end{equation}

We conclude that, for $\eta_{i}\to-\infty$ (when all the modes are inside the Hubble radius initially), there is only one minimum for this integral, and therefore, only one vacuum state, correspondingly. Moreover, this vacuum state coincides with the already mentioned privileged de Sitter vacuum that is invariant under spatial translations and rotations, dilations of space and time, and such that its 2-point function fulfills the Hadamard condition (i.e. the Bunch-Davies state).

\end{document}